\documentclass[aps, prd, twocolumn, 
showpacs, showkeys, 
print, preprintnumbers, nofootinbib,
superscriptaddress 
]{revtex4-2}
\usepackage{feynmf}
\usepackage{soul}
\usepackage{mciteplus}
\usepackage[utf8]{inputenc}
\usepackage[T1]{fontenc}
\usepackage{lmodern}
\usepackage{microtype}
\usepackage[english]{babel}
\usepackage{amssymb,amsbsy,amsmath,amsfonts,amsthm}
\usepackage{yhmath} 
\usepackage{slashed}
\usepackage{bm}
\usepackage{times}
\usepackage[dvipsnames,table,xcdraw]{xcolor}
\definecolor{lightyellow}{RGB}{255,250,205}
\usepackage{graphicx}
\usepackage{epstopdf}
\usepackage{tabularx}
\usepackage{multirow}
\usepackage{longtable}
\usepackage{hyperref}
\hypersetup{colorlinks=true, linkcolor=blue, 
citecolor=cyan, urlcolor=black}
\usepackage{url}
\newcommand{\mathleft}{\@fleqntrue\@mathmargin0pt}
\newcommand{\mathcenter}{\@fleqnfalse}
\usepackage{soul}
\usepackage{orcidlink}
\allowdisplaybreaks

\begin{document}

\title{Dalitz-plot decomposition for the $e^+ e^- \to J/\psi \, \pi \, \pi \, (K \bar{K})$ and $e^+ e^- \to h_c \, \pi \, \pi$ processes}
\author{Viktoriia Ermolina\orcidlink{0009-0005-6965-4840}}
\email[]{vermolin@uni-mainz.de}
\affiliation{Institut f\"ur Kernphysik \& PRISMA$^+$ Cluster of Excellence, Johannes Gutenberg Universit\"at,  D-55099 Mainz, Germany}

\author{Igor Danilkin\orcidlink{0000-0001-8950-0770}}
\affiliation{Institut f\"ur Kernphysik \& PRISMA$^+$ Cluster of Excellence, Johannes Gutenberg Universit\"at,  D-55099 Mainz, Germany}

\author{Marc Vanderhaeghen\orcidlink{0000-0003-2363-5124}}
\affiliation{Institut f\"ur Kernphysik \& PRISMA$^+$ Cluster of Excellence, Johannes Gutenberg Universit\"at,  D-55099 Mainz, Germany}

\begin{abstract}
We present an analysis of the $e^+ e^- \to \gamma^* \to J/\psi \, \pi \, \pi \, (K \bar{K})$ and $e^+ e^- \to \gamma^* \to h_c \, \pi \, \pi$ processes employing the recently proposed Dalitz-plot decomposition approach, which is based on the helicity formalism for three-body decays. For the above reactions, we validate the factorization of the overall rotation for all decay chains and spin alignments, along with the crossing symmetry between final states, using a Lagrangian-based toy model.  For the model-dependent factors that describe the subchannel dynamics, we employ the dispersive treatment of the $\pi \, \pi \, (K \bar{K})$ final state interaction, which accurately reproduces pole positions and couplings of the $f_0(500)$ and $f_0(980)$ resonances. The constructed amplitudes serve as an essential framework to further constrain the properties of the charged exotic states $Z_c(3900)$ and $Z_c(4020)$, produced in these reactions.

\end{abstract}
\maketitle
\date{\today}

\section{Introduction}\label{introduction}

The understanding of the spectrum and quantum numbers of hadrons, which apart from a few exceptions are unstable due to strong interactions, requires a formalism to describe their decay. The analysis of decay amplitudes to final states with various non-zero spins has been widely examined, including both covariant \cite{Rarita:1941mf,Zemach:1963bc} and non-covariant methods \cite{Zemach:1965ycj}. An example of the latter is the helicity formalism, originally developed by Jacob and Wick \cite{Jacob:1959at}, which established a framework for studying sequential decays \cite{Chung:2007nn}. This formalism is a powerful tool for isolating the contributions of particular spin and parity $J^P$, which allows for the determination of the quantum numbers of newly discovered resonances. Furthermore, this approach enables the study of processes involving several decay chains leading to the same final states, provided that their helicities are aligned in the same reference frame to perform the summation over them \cite{Chen:2017gtx}. The helicity formalism approach takes into account distinct contributions from angular distributions using Wigner $D$-functions and kinematic dependence. For the case of three particles in the final state, on which we focus in this work, the full amplitude includes the sum over the three different decay chains with a certain set of angles entering each of them. However, by employing the factorization method of the Dalitz-plot decomposition (DPD) proposed in \cite{JPAC:2019ufm}, it becomes possible to isolate the decay plane's orientation and express the remaining $D$-function arguments in terms of Mandelstam variables. It results into the full amplitude being simplified significantly and depending only on five variables: three Euler angles which set the orientation of decay plane and two Mandelstam variables involved in the Dalitz-plot function. The DPD formalism also allows for extension to cascade reactions \cite{Habermann:2024sxs}.

In the present paper, we focus on $e^+e^-$ reactions, which serve as an important probe in the search for new resonances at BESIII and Belle II collider experiments. In recent years, a plethora of new states containing heavy charm and bottom quarks have been found, which could not be understood as bound states of a quark and antiquark, and require more exotic treatment as tetraquark or molecular states \cite{Brambilla:2019esw,Chen:2016qju,Chen:2022asf,Olsen:2017bmm,Karliner:2017qhf,Guo:2017jvc,Liu:2019zoy}. In the analyses of $ e^+ e^- \to J/\psi \, \pi^+ \pi^- \, (K\bar{K}), \ h_c \, \pi^+ \pi^- $ with the DPD, the crossing symmetry of the pions results in a reduction in the number of independent helicity couplings, but requires the inclusion of additional phase factors. Since the separation of the angular variables from the dynamical ones is model-independent, in this paper we show the validation of the results of the DPD irrespective of the selected reference frame using a toy-model Lagrangian, which is undertaken for the first time.

The system of two pions, which enters the reactions under study in the present paper, frequently appears as a part of the final state in many hadronic interactions, making it an essential input in various analyses of experimental data. The $\pi\pi \to \pi\pi$ and $\pi\pi \to K\bar{K}$ amplitudes are well-known from the Roy (Roy-Steiner) analyses \cite{Garcia-Martin:2011iqs,Buettiker:2003pp,Pelaez:2018qny}, which incorporate all the fundamental S-matrix constraints: unitarity, analyticity, and crossing symmetry. Nevertheless, these amplitudes cannot be directly applied in experimental data analyses due to different left-hand cuts for each production or decay mechanism. Since unitarity is the main principle that provides a connection between the production/decay and scattering amplitude, the correct implementation of the $\pi\pi$ rescattering must be performed using the so-called Omnès matrix, which only has right-hand cuts. In practice, however, $\pi\pi$ final state interactions (FSI) are typically described phenomenologically, either as a sum of Breit-Wigner amplitudes \cite{Pilloni:2016obd} or as e.g. a combination of Breit-Wigner for the $f_0(500)$ and Flatté for the $f_0(980)$ \cite{BESIII:2017bua,BESIII:2020oph}. Both approaches violate unitarity, and the resonance parameters do not accurately reflect the $\pi\pi/K\bar{K}$ phase shifts. Progress has been achieved in \cite{Danilkin:2020kce}, where it was shown that the $\pi\pi$ mass distribution of $e^+e^- \to J/\psi \, \pi \, \pi \, (K\bar{K})$ can be efficiently described by the Omnès matrix \cite{Danilkin:2020pak} multiplied by a subtraction polynomial. 

In this paper, we demonstrate how to incorporate this result into the model-dependent part of the DPD. In contrast to the \cite{Danilkin:2020kce}, the approach discussed in the present paper allows 
to consider different quantum numbers of a resonance in a straightforward way and obtain its angular dependence. Having the shape of angular distributions is imperative to determine the quantum numbers of an observed resonance from comparison to the experimental data. Using this approach, we perform a test fit to the available empirical data on invariant mass distributions of the $e^+e^- \to J/\psi \, \pi \, \pi \, (K\bar{K})$. 

We begin by discussing the DPD formalism in Sec. \ref{2}. In Sec. \ref{3}, we demonstrate examples of helicity amplitudes in the two specific cases of three-body decays, $e^+ e^- \to J/\psi \, \pi^+ \, \pi^- \, (K^+ K^-)$ and $e^+ e^- \to h_c \, \pi^+ \, \pi^-$, which involve axial-vector, scalar, or tensor resonances. The cross-validation with the toy-model Lagrangian description is  discussed in details. We provide a coupled-channel dispersive model of the $\pi\pi(K\bar{K})$ final state interaction in Sec. \ref{4}. Within this approach in Sec. \ref{5}, we fit the experimental invariant mass distributions measured by the BESIII Collaboration \cite{BESIII:2017bua,BESIII:2022joj} at different $e^+e^-$ CM energies.

\section{Formalism}\label{2}

To describe the Dalitz-plot decomposition of $e^+ e^- \to J/\psi \, \pi^+ \, \pi^- $ and related processes, we start by briefly reviewing the basic definitions of \cite{JPAC:2019ufm}. For a three-body decay $0\to 123$, the transition amplitude can be written as
\begin{equation}\label{eq:0}
    M^\Lambda_{\{\lambda\}} =\sum_{\nu} D^{J*}_{\Lambda,\nu}(\phi_1,\theta_1,\phi_{23})\,O^\nu_{\{\lambda\}}(\{\sigma\})\,,
\end{equation}
where the particle in the initial state has spin $J$ and spin projection $\Lambda$ quantized along the $z$ axis. Individual helicities of final states are collectively labeled as ${\{\lambda\}}\equiv (\lambda_1,\lambda_2,\lambda_3)$. The rotation with Wigner $D$-function connects the CM frame of calculation (with the decay-product plane chosen to be the $xz$ plane with the momentum $-\vec{p}_1$ directed along the $z$ axis)\footnote{Within the DPD formalism, the convention to factor out the D-function of the particle-1 is assumed. Determining which particles are to be designated as 1, 2, and 3 is an arbitrary choice.} to the actual CM frame of reference, whose position in space is defined by Euler angles $(\phi_1,\theta_1,\phi_{23})$.

In this construction (\ref{eq:0}), the angular variables are separated in a model independent way from the dynamical variables 
\begin{equation}\label{eq:sigma_i}
    \sigma_1=(p_2+p_3)^2, \ \ \sigma_2=(p_1+p_3)^2, \ \ \sigma_3=(p_1+p_2)^2\,,
\end{equation}
which enter the Dalitz-plot function $O^\nu_{\{\lambda\}}$. The latter is given by a product of individual two-particle decays, each one considered in the rest frame of a decaying particle
\begin{equation}\label{eq:DPD_O_general}
\begin{split}
    O^\nu_{\{\lambda\}}(\{\sigma\}) & = \sum_{(ij)k} \sum_s^{(ij)\to i,j} \sum_\tau \sum_{\{\lambda'\}} n_J\, n_s\,d^J_{\nu,\tau-\lambda'_k} (\hat{\theta}_{k(1)}) \\& \times \,H^{0 \to (ij),k}_{\tau,\lambda'_k}X_s (\sigma_k)\,d^s_{\tau,\lambda'_i-\lambda'_j}(\theta_{ij})\,H^{(ij) \to i,j}_{\lambda'_i,\lambda'_j} \\& \times d^{j_1}_{\lambda'_1,\lambda_1} (\zeta^1_{k(0)})\, d^{j_2}_{\lambda'_2,\lambda_2} (\zeta^2_{k(0)})\, d^{j_3}_{\lambda'_3,\lambda_3} (\zeta^3_{k(0)})\,.
\end{split}
\end{equation}
The rotation by the angles $\hat{\theta}_{k(1)}$ relates all the three chains with each other, which is achieved by choosing the specific frame of calculation, while the angles $\theta_{ij}$ denote the polar angle of particle-$i$ in $(ij)$ rest frame. Finally, a boost that induces an additional rotation of helicities corresponding to each final state by the angles $\zeta^{1,2,3}_{k(0)}$ connects the individual two-particle decays. The detailed expressions for the angles $\hat{\theta}_{k(1)},\theta_{ij},\zeta^{1,2,3}_{k(0)}$, in terms of $\sigma_{1,2,3}$, can be found in Appendix A of \cite{JPAC:2019ufm}.

In Eq. (\ref{eq:DPD_O_general}) the first sum is taken over all possible configurations $(ij)k \in \{(23)1,(31)2,(12)3\}$, which correspond to three potential decay chains, the second and third summation is over various possible spins $s$ and helicity $\tau$ of isobar $(ij)$. Furthermore, the functions $H$ denote helicity couplings, the functions $X_s (\sigma)$ specify the energy dependence of the isobar, $n_J$ and $n_s$ serve as conventional normalization factors. Individual spins of final states are denoted as $j_i$. If the parity is conserved, the helicity couplings are related as 
\begin{align}
    &H^{0 \to (ij),k}_{\tau,\lambda'_k}=(-1)^{-J+s+j_k} P_{0}\,P_{(ij)}\,P_k \,H^{0 \to (ij),k}_{-\tau,-\lambda'_k},\\
    &H^{(ij) \to i,j}_{\lambda'_i,\lambda'_j}=(-1)^{-s+j_i+j_j} P_{(ij)}\,P_i\,P_j \,H^{(ij) \to i,j}_{-\lambda'_i,-\lambda'_j},
\end{align}
with $P_n$ standing for the intrinsic parity of the corresponding particle. The number of independent couplings $H$ can be further reduced when there is a permutation symmetry between two or all three final states. To fit the generally unknown couplings $H$ to the available data, it is common to employ the $LS$ helicity coupling scheme \cite{Chung:1997jn}. For the particle $0$, decaying into an isobar $(ij)$ and a final state $k$, as well as for the isobar $(ij)$ with its decay products $i,j$, the decompositions are given by
\begin{equation}\label{eq:ls_decomposition}
\begin{split}
        &H^{0 \to (ij),k}_{\tau,\lambda'_k}\begin{aligned}[t]&=\sum_{LS} \alpha^{0 \to (ij),k}_{LS} \sqrt{\frac{2L+1}{2J+1}} \langle s,\tau;j_k,-\lambda'_k|S,\tau-\lambda'_k \rangle \\
        &\times \langle L,0;S,\tau-\lambda'_k|J,\tau-\lambda'_k \rangle p^L \,B_L\,,
        \end{aligned} 
        \\
        &H^{(ij) \to i,j}_{\lambda'_i,\lambda'_j}\begin{aligned}[t]&=\sum_{l's'} \alpha^{(ij) \to i,j}_{l's'} \sqrt{\frac{2l'+1}{2s+1}} \langle j_i,\lambda'_i;j_j,-\lambda'_j|s',\lambda'_i-\lambda'_j \rangle \\
        &\times \langle l',0;s',\lambda'_i-\lambda'_j|s,\lambda'_i-\lambda'_j \rangle {p'}^{l'}\, B_{l'}\,.
        \end{aligned}
\end{split}
\end{equation}
Here $\alpha^{0 \to (ij),k}_{LS}$ and $\alpha^{(ij) \to i,j}_{l's'}$ stand for $LS$ couplings of the corresponding decay, $S$ denotes the spin of the isobar-spectator system, $s'$ is the spin of the $i$-$j$ system, and $L, l'$ are the relative orbital angular momenta between final particles. The spins of decaying particle $0$ and isobar $(ij)$ are $J$ and $s$, respectively. The magnitude of $\vec{p}_k$ or $\vec{p}_i+\vec{p}_j$ in the rest system frame of particle $0$ is denoted by $p$ \cite{Zou:2002ar}, $p'$ is the magnitude of $\vec{p}_i$ or $\vec{p}_j$, while $B_L, B_{l'}$ are the Blatt-Weisskopf functions (normalized to $1$ at the resonance position) \cite{VonHippel:1972fg}, which guarantee the proper asymptotic behavior. 

\section{Application to $ e^+ e^- \to J/\psi \, \pi \, \pi \, (K \bar{K}), \ h_c \, \pi \, \pi$ processes}\label{3}

We now apply the DPD to the processes $e^+ e^- \to J/\psi \, \pi \, \pi \, (K\bar{K})$ and $e^+ e^- \to h_c \, \pi \, \pi$, which serve as discovery channels for the $Z_c(3900)$ \cite{Belle:2013yex,BESIII:2013mhi} and $Z_c(4020)$ \cite{BESIII:2013ouc}, respectively. These processes exhibit specific properties related to the crossing symmetry of pions and particular helicities of the intermediate virtual photon. In the following, we start with applying the DPD to the three-body decay cases $\gamma^* \to J/\psi \, \pi \, \pi \, (K\bar{K}), \, h_c \, \pi \, \pi$, and then extend the results to the $2 \to 3$ processes, when the virtual photon is produced in the $e^+ e^-$ collision.

\subsection{Unpolarized three-body decay case}

For the process $\gamma^* \to J/\psi \, \pi^+ \, \pi^-$, each of the three decay chains has at least one resonance in the Dalitz-plot decomposition, with $Z^{\pm}_c(3900)$ in $J/\psi \, \pi^{\pm}$ and $f_0(500), f_0(980), f_2(1270)$ in $\pi^+\pi^-$ \cite{BESIII:2017bua}. We denote the particles as $1=J/\psi, \, 2=\pi^-, \, 3=\pi^+$, which enables the symmetric form of $Z^{\pm}_c(3900)$ contributions to a matrix element. The Dalitz-plot function reads 
\begin{widetext}

\begin{flalign}\label{eq:2}
\begin{split}
\includegraphics[width=0.25\textwidth]{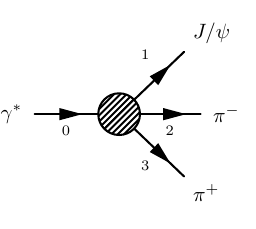}
\end{split}
\begin{split}
O^\nu_{\lambda_1}(\{\sigma\}) & = n_1 n_0 \delta_{\nu,-\lambda_1} \left( \mathring{H}^{0 \to (23),1}_{0,\lambda_1}(\sigma_1) \mathring{X}_0 (\sigma_1) \mathring{H}^{(23) \to 2,3}_{0,0} + \dot{H}^{0 \to (23),1}_{0,\lambda_1}(\sigma_1) \dot{X}_0 (\sigma_1) \dot{H}^{(23) \to 2,3}_{0,0} \right)\\& +
\sum_\tau n_1 n_2 \delta_{\nu,\tau-\lambda_1} \breve{H}^{0 \to (23),1}_{\tau,\lambda_1}(\sigma_1) \breve{X}_2 (\sigma_1) d^2_{\tau,0} (\theta_{23}) \breve{H}^{(23) \to 2,3}_{0,0}(\sigma_1) \\
& + \sum_{\tau,\lambda'_1} n_1^2 d^1_{\nu,\tau} (\hat{\theta}_{2(1)}) H^{0 \to (31),2}_{\tau,0}(\sigma_2) X_1 (\sigma_2) d^1_{\tau,-\lambda'_1} (\theta_{31}) H^{(31) \to 3,1}_{0,\lambda'_1}(\sigma_2) d^{1}_{\lambda'_1,\lambda_1} (\zeta^1_{2(0)})\\
& + \sum_{\tau,\lambda'_1} n_1^2 d^1_{\nu,\tau} (\hat{\theta}_{3(1)}) H^{0 \to (12),3}_{\tau,0}(\sigma_3) X_1 (\sigma_3) d^1_{\tau,\lambda'_1} (\theta_{12}) H^{(12) \to 1,2}_{\lambda'_1,0}(\sigma_3) d^{1}_{\lambda'_1,\lambda_1} (\zeta^1_{3(0)}),
\end{split}
\end{flalign}
\end{widetext}
where the first row accounts for scalar resonances $f_0(500)$ with couplings $\mathring{H}$ and functions $\mathring{X}$, and $f_0(980)$ with couplings $\dot{H}$ and functions $\dot{X}$, the subsequent - for tensor resonance $f_2(1270)$ with couplings $\breve{H}$ and functions $\breve{X}$, and the last two describe exotic resonance $Z^{\pm}_c(3900)$ with couplings $H$ and functions $X$. The normalization constants are defined as $n_l=\sqrt{(2l+1)/4\pi}$. Following \cite{BESIII:2017bua}, we consider the $Z^{\pm}_c(3900)$ state to be an axial vector $1^+$. However, this methodology permits any possible configuration of quantum numbers to be taken into consideration, or to be identified through fitting to experimental data. 

For an unpolarized three-body decay, the cross section is fully determined by the dynamical variables of the matrix element and proportional to the squared of the Dalitz-plot function $O^{\nu}_{\lambda_1}(\{\sigma\})$
\begin{equation}\label{eq:cross-check-1}
\frac{d\sigma}{d\sigma_1\,d\sigma_2} \sim \displaystyle\sum_{\lambda_1,\Lambda} \left|M^{\Lambda}_{\lambda_1} \right|^2=\displaystyle\sum_{\lambda_1,\nu} \left|O^{\nu}_{\lambda_1}(\{\sigma\}) \right|^2.
\end{equation}
Since the separation of the angular variables from the dynamical ones in Eq. (\ref{eq:2}) is model-independent, we decided to validate the results of the DPD using the toy-model Lagrangian, which includes all relevant vertices with coupling constants fixed to one. The explicit form of the Lagrangian is given in Appendix \ref{L}. The calculations of (\ref{eq:cross-check-1}) can be carried out in two ways. The first one is directly from the Lagrangian by employing the completeness relation for the polarization vectors, thus being independent of their particular form. The second way is through the DPD given in Eq. (\ref{eq:2}), where the helicity couplings $H$ are extracted from the corresponding two-particle decays using the same Lagrangian and the standard helicity formalism. 
For example, the transition amplitude for the two-body decay $(ij) \to i j$ in the rest frame of $(ij)$ is given by \cite{Chung:1971ri}
\begin{equation}
A^{s,\tau}_{\lambda_i,\lambda_j} = n_{s}\, H^{(ij) \to i,j}_{\lambda_i,\lambda_j}(\sigma_k)\, D^{s*}_{\tau,\lambda_i-\lambda_j} (\phi_{ij},\theta_{ij},0)\,,
\end{equation}
where $(\theta_{ij},\phi_{ij})$ represent the direction of the momentum of particle $i$. The functions $X(\{\sigma\})$ are simply scalar propagators. Since the considered resonances are in the physical region of the Dalitz plot, to obtain finite results in both calculations, we introduced a constant width in the propagators using the simple prescription $m_R \to m_R - i\, \Gamma_R/2$. When calculating the helicity couplings for the $Z_c^{\pm}$ contributions, one needs to be careful that the crossing symmetry is not naturally implemented in Eq. 7 due to cyclic permutations. Therefore, one needs to account for an additional phase factors \cite{JPAC:2019ufm, JPAC:2021rxu}
\begin{align}\label{eq:crossing_relations}
& H^{(31) \to 3,1}_{0,\lambda'_1}(\sigma_2)=(-1)^{1-\lambda'_1}\,H^{(12) \to 1,2}_{\lambda'_1,0}(\sigma_3 \to \sigma_2)\,,\nonumber\\
& H^{0\to (31),2}_{\tau,0}(\sigma_2)=H^{0 \to (12),3}_{\tau,0}(\sigma_3 \to \sigma_2)\,.
\end{align}
These relations allow to reduce the amount of independent helicity couplings and consequently, the number of unknown $LS$ couplings $\alpha$ in Eq. (\ref{eq:ls_decomposition}). Conversely, without these additional phase factors, one can not use the same $LS$ couplings for the $Z_c^+$ and $Z_c^-$ contributions, as their relations with each other do not follow from the definition of helicity couplings. By comparing two calculations using the same Lagrangian, we confirm the correctness of the DPD and validate three important points: the factorization of the overall rotation, the spin alignments, and the crossing symmetry between the two final states. 

To complete the cross-verification,  we compare the results obtained within a different set of particles configuration. The latter ensures the Lorenz invariance of the approach. For that purpose, we denote the particles as $1=\pi^-, \, 2=J/\psi, \, 3=\pi^+$. The Dalitz-plot function in this case reads 
\begin{widetext}

\begin{flalign}\label{eq:3}
\begin{split}
\includegraphics[width=0.25\textwidth]{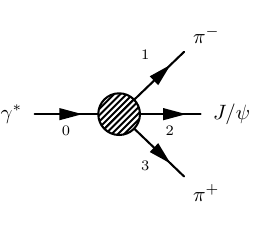}
\end{split}
\begin{split}
O^\nu_{\lambda_2}(\{\sigma\}) & = n_1 n_0 d^1_{\nu,-\lambda_2}(\hat{\theta}_{2(1)}) \left( \mathring{H}^{0 \to (31),2}_{0,\lambda_2} \mathring{X}_0 (\sigma_2) \mathring{H}^{(31) \to 3,1}_{0,0} +\dot{H}^{0 \to (31),2}_{0,\lambda_2} \dot{X}_0 (\sigma_2) \dot{H}^{(31) \to 3,1}_{0,0} \right)\\
& +\sum_\tau n_1 n_2 d^1_{\nu,\tau-\lambda_2} (\hat{\theta}_{2(1)}) \breve{H}^{0 \to (31),2}_{\tau,\lambda_2} \breve{X}_2 (\sigma_2) d^2_{\tau,0} (\theta_{31}) \breve{H}^{(31) \to 3,1}_{0,0} \\
& + \sum_{\tau,\lambda'_2} n_1^2 \delta_{\nu,\tau} H^{0 \to (23),1}_{\tau,0} X_1 (\sigma_1) d^1_{\tau,\lambda'_2} (\theta_{23}) H^{(23) \to 2,3}_{\lambda'_2,0} d^{1}_{\lambda'_2,\lambda_2} (\zeta^2_{1(0)})\\
& + \sum_{\tau,\lambda'_2} n_1^2 d^1_{\nu,\tau} (\hat{\theta}_{3(1)}) H^{0 \to (12),3}_{\tau,0} X_1 (\sigma_3) d^1_{\tau,-\lambda'_2} (\theta_{12}) H^{(12) \to 1,2}_{0,\lambda'_2} d^{1}_{\lambda'_2,\lambda_2} (\zeta^2_{3(0)}).
\end{split}
\end{flalign}

\noindent Note that, for brevity, we distinguish Eq. (\ref{eq:2}) and Eq. (\ref{eq:3}) by the helicity label of the $J/\psi$ (i.e. in Eq. (\ref{eq:2}) $J/\psi$ is particle-1, while in Eq. (\ref{eq:3}) it is particle-2). Analogous expression can be obtained within the $1=\pi^-, \, 2=\pi^+, \, 3=J/\psi$ configuration, which we denote as $O^{\nu}_{\lambda_3}$. The validation of the DPD renders
\begin{equation}
    \sum_{\lambda_1,\nu} \left|O^{\nu}_{\lambda_1} \right|^2=\sum_{\lambda_2,\nu} \left|O^{\nu}_{\lambda_2} \right|^2=\sum_{\lambda_3,\nu} \left|O^{\nu}_{\lambda_3} \right|^2.
\end{equation}
The process $\gamma^* \to J/\psi \, K^+ K^-$ is analogous (the same quantum numbers are involved).

The next example is $\gamma^* \to h_c \, \pi \, \pi$. This process serves as a discovery channel for the $Z^{\pm}_c(4020)$ state \cite{BESIII:2013ouc}, while it may contain the aforementioned state $Z^{\pm}_c(3900)$ as well in the physical region. Therefore, we assume the presence of $Z^{\pm}_c(3900)$ and $Z^{\pm}_c(4020)$ in $h_c \pi^{\pm}$, and $f_0(500)$ in $\pi^+\pi^-$. Compared to the previous example, this process is distinguished by the parity of the final charmonium $J^P(h_c) = 1^+$. The latter affects only the form of the helicity couplings and the allowed $LS$ $(l's')$ combinations, which are provided in Table \ref{Tableq1}. We denote the particles as $1 = \pi^-,\, 2 = h_c,\, 3 = \pi^+$. In this setup, $\theta_1$ defines the polar angle of $Z_c^+$. This will be important for the $2 \to 3$ case, as it allows us to extract the angular dependence and ultimately determine the spin and parity of $Z_c^+$. The Dalitz-plot function has the following form 

\begin{flalign}\label{eq:DPD_hc}
\begin{split}
\includegraphics[width=0.25\textwidth]{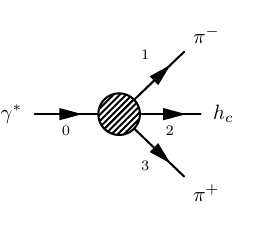}
\end{split} \begin{split}
O^\nu_{\lambda_2}(\{\sigma\}) & = n_1 n_0 d^1_{\nu,-\lambda_2} (\hat{\theta}_{2(1)}) \mathring{H}^{0 \to (31),2}_{0,\lambda_2} \mathring{X}_0 (\sigma_2) \mathring{H}^{(31) \to 3,1}_{0,0} \\
& + \sum_{\tau,\lambda'_2} n_1\delta_{\nu,\tau}\left(n_1  H^{0 \to (23),1}_{\tau,0} X_1 (\sigma_1) d^1_{\tau,\lambda'_2} (\theta_{23}) H^{(23) \to 2,3}_{\lambda'_2,0}  \right.\\
&\hspace{2cm} \left.+\,n_s  \bar{H}^{0 \to (23),1}_{\tau,0} \bar{X}_s (\sigma_1) d^s_{\tau,\lambda'_2} (\theta_{23}) \bar{H}^{(23) \to 2,3}_{\lambda'_2,0}\right) d^{1}_{\lambda'_2,\lambda_2} (\zeta^2_{1(0)}) \\
&+\sum_{\tau,\lambda'_2} n_1  d^1_{\nu,\tau}(\hat{\theta}_{3(1)}) \left(n_1 H^{0 \to (12),3}_{\tau,0}  X_1 (\sigma_3) d^1_{\tau,-\lambda'_2} (\theta_{12}) H^{(12) \to 1,2}_{0,\lambda'_2} \right.\\
&\hspace{2cm} \left.+\,n_s \bar{H}^{0 \to (12),3}_{\tau,0}  \bar{X}_s (\sigma_3) d^s_{\tau,-\lambda'_2} (\theta_{12}) \bar{H}^{(12) \to 1,2}_{0,\lambda'_2}\right) d^{1}_{\lambda'_2,\lambda_2} (\zeta^2_{3(0)}),
  \end{split}
\end{flalign}
\end{widetext}
where the first row accounts for scalar resonance $f_0(500)$, followed by the ones for the $Z^{\pm}_c(3900)$ and $Z^{\pm}_c(4020)$ resonances, the latter with couplings $\bar{H}$ and functions $\bar{X}$ having an arbitrary spin $s$. For the purpose of the verification of DPD by the Lagrangian we assumed all $Z_c$ to be axial vectors. Using vector-axial-scalar and scalar-pseudoscalar-pseudoscalar vertices from Appendix \ref{L}, we recover the same expressions for the matrix element squared by applying the algorithm discussed in the previous example.

\subsection{Unpolarized $e^+ e^- \to 123$ case}

For the full $e^+ e^- \to 123$ process the unpolarized differential cross-section for is given by 
\begin{align}\label{eq:cross_section_polarized}
   & \frac{d\sigma}{d\cos{\theta_1} \, d\phi_{23} \, ds \, dt}=\frac{e^2}{64\,(2\pi)^4\,q^6}  \\
   & \quad\quad\quad \times \sum_{\{\lambda\}} \left(\left| M^{\Lambda=+1}_{\{\lambda\}} \right|^2+\frac{2m^2_e}{q^2}\left| M^{\Lambda=0}_{\{\lambda\}} \right|^2\right),\nonumber
\end{align}
As one notices, the contribution of the helicity amplitude associated with the helicity zero of the photon to the differential cross-section is suppressed by a factor of $2m^2_e/q^2$, with $m_e$ being electron mass and $q$ representing the $e^+e^-$ CM energy. Consequently, for $q^2 \gg m_e^2$ one can neglect the term with longitudinal polarization. The formula (\ref{eq:cross_section_polarized}) allows for the determination of the polar angular distribution of the isobar $(ij)=(23)$. This approach ensures that the total cross-sections are equal for the three possible configurations, where $J/\psi$ (or $h_c$) is denoted as particle-1 ($O^{\nu}_{\lambda_1}$), particle-2 ($O^{\nu}_{\lambda_2}$), or particle-3 ($O^{\nu}_{\lambda_3}$), as introduced in Eqs. (\ref{eq:2}) and (\ref{eq:3}). Additionally, it was verified that the matrix element obtained using the helicity formalism framework in Eq. (\ref{eq:cross_section_polarized}), with helicity coupling coefficients derived from the Lagrangian, agrees with the straightforward Lagrangian calculation for all three configurations.

\section{Dispersive treatment of final state interactions}\label{4}

The product of functions of a single Mandelstam variable, $H^{0\to(ij)k}\,X\,H^{(ij)\to i,j}$, is the only model-dependent component of the DPD. Its dispersive treatment using the Khuri-Treiman approach \cite{Khuri:1960zz} is typically numerically demanding and requires detailed knowledge of the all two-body phase shifts. Additionally, extending this method to coupled channels or particles with arbitrary spins is cumbersome and rarely used in practical applications \cite{Albaladejo:2017hhj, Albaladejo:2019huw, Stamen:2022eda}. However, this does not imply that some aspects of the overall problem cannot be treated more accurately than in the Breit-Wigner or Flattè approximations. Below, we focus on the S-wave, which has no complications related to the kinematic constraints, mainly arising from the threshold factors.

The isobar model is nothing else but a truncated partial wave expansion that provides an effective description of intermediate resonances. The key distinction between the Breit-Wigner parameterization and the dispersive treatment lies in the unitarity constraint imposed in all three channels. Its implementation on a truncated set leads to two contributions: dominant two-body rescattering and so-called three-body (crossed-channel) effects. Since the main mechanism remains two-body rescattering, the final result is governed by the standard Omnès formalism. The contributions from crossed-channel rescattering, which correspond to the left-hand cuts, can be absorbed in the subtraction polynomial, as we have shown in \cite{Danilkin:2020kce}. In the single-channel case (which considers only the $f_0(500)$ resonance), which is relevant for processes like $e^+e^- \to h_c \, \pi \, \pi$, this corresponds to the replacement in Eq. (\ref{eq:DPD_hc}):
\begin{equation}\label{eq:FSI_pipi_1}
\mathring{\alpha}^{0 \to (31),2}_{11} \mathring{X}_0 (\sigma_2)\, \mathring{\alpha}^{(31) \to 3,1}_{00}  =(a+b\,\sigma_2)\,\Omega(\sigma_2),
\end{equation}
with $a$ and $b$ being unknown real parameters and
\begin{equation}\label{Eq:SCOmnes}
\Omega(\sigma)=\exp\left(\frac{\sigma}{\pi}\int_{4m^2_\pi}^\infty \frac{d\sigma'}{\sigma'}\frac{\delta(\sigma')}{\sigma'-\sigma} \right).
\end{equation}
In (\ref{Eq:SCOmnes}), the single-channel $\pi\pi$ $S$-wave isospin $I=0$ phase shift can be taken from different dispersive analyses \cite{Danilkin:2020pak, Danilkin:2022cnj}, which give very similar results. For instance, the phase-shift from \cite{Danilkin:2020pak} corresponds to the pole on RSII
\begin{equation}
    \sqrt{s_{f_0(500)}}=458(7)^{+4}_{-10} - i\, 245(6)^{+7}_{-10} \text{ MeV},
\end{equation}
in agreement with \cite{Pelaez:2015qba}. If the kinematical region extends beyond the inelastic $K\bar{K}$ channel, one must use the coupled-channel approach, which accounts for both $f_0(500)$ and $f_0(980)$. This is relevant, for example, in processes like $e^+e^- \to \pi^+\pi^- J/\psi$, and corresponds to a similar replacement in Eq. (\ref{eq:2})
\begin{align} \label{eq:FSI_pipi_2}
&\mathring{\alpha}^{0 \to (23),1}_{01} \mathring{X}_0 (\sigma_1) \, \mathring{\alpha}^{(23) \to 2,3}_{00}+ \dot{\alpha}^{0 \to (23),1}_{01} \dot{X}_0 (\sigma_1) \, \dot{\alpha}^{(23) \to 2,3}_{00} \nonumber \\
& =(a+b\,\sigma_1)\,\Omega^{(0)}_{11}(\sigma_1) + (c+d\,\sigma_1)\,\Omega^{(0)}_{12}(\sigma_1),
\end{align}
while for the $S$-wave term in $e^+e^- \to K^+K^- J/\psi$ one needs to use
\begin{equation} \label{eq:FSI_KK}
    \frac{\sqrt{3}}{2}\left((a+b\,\sigma_1)\,\Omega^{(0)}_{21}(\sigma_1) + (c+d\,\sigma_1)\,\Omega^{(0)}_{22}(\sigma_1)\right),
\end{equation}
which contains an additional $\sqrt{3}/2$ factor due to isospin. The coupled-channel Omnès matrix we propose is based on a data-driven $N/D$ analysis \cite{Danilkin:2020pak}, where the fit is performed using the latest Roy and Roy–Steiner results for $\pi\pi \to \pi\pi$ \cite{Garcia-Martin:2011iqs} and $\pi\pi \to \bar{K}K$ \cite{Pelaez:2020gnd}, respectively. Through analytic continuation into the complex plane, this solution yields poles for the $f_0(500)$ and $f_0(980)$ at 
\begin{equation}
\begin{split}
     &\sqrt{s_{f_0(500)}}=458(10)^{+7}_{-15} - i\, 256(9)^{+5}_{-8} \text{ MeV}, \\
     &\sqrt{s_{f_0(980)}}=993(2)^{+2}_{-1} - i\,21(3)^{+2}_{-4} \text{ MeV},
\end{split}
\end{equation}
which are in good agreement with Refs. \cite{Caprini:2005zr, Garcia-Martin:2011nna, Moussallam:2011zg}. For other implementations of the $\pi\pi/K\bar{K}$ Omnès matrix, see \cite{Donoghue:1990xh, Moussallam:1999aq, Hoferichter:2012wf}, or more recent works such as \cite{Blackstone:2024ouf, Pich:2023kim}.

\begin{table}[t!]
\centering 
\renewcommand*{\arraystretch}{1.4}
\begin{tabular*}{\columnwidth}{@{\extracolsep{\fill}}lc}
  \hline
 Decay & Corresponding $LS$ $(l's')$ combinations \\ 
 \hline
\multicolumn{2}{c}{$e^+ e^- \to J/\psi \, \pi^+ \, \pi^-$}\\
 \hline
 $\gamma^* \to Z^\pm_c\pi^\mp$ & $(0,1)$, $(2,1)$ \\ 
 $Z^\pm_c \to J/\psi\pi^\pm$ & $(0,1)$, $(2,1)$ \\ 
 $\gamma^* \to f_0 J/\psi$ & $(0,1)$, $(2,1)$ \\
 $f_0 \to \pi^+\pi^-$ & $(0,0)$ \\
 $\gamma^* \to f_2 J/\psi$ & $(0,1)$, $(2,1)$, $(2,2)$, $(2,3)$, $(4,3)$ \\
 $f_2 \to \pi^+ \pi^-$ & $(2,0)$\\
 \hline
 \multicolumn{2}{c}{$e^+ e^- \to h_c \, \pi^+ \, \pi^-$}\\
  \hline
 $\gamma^* \to Z^\pm_c \pi^\mp$ & $(0,1)$, $(2,1)$ \\ 
 $Z^\pm_c \to h_c \pi^\pm$ & $(1,1)$ \\ 
 $\gamma^* \to f_0 h_c$ & $(1,1)$ \\
 $f_0 \to \pi^+ \pi^-$ & $(0,0)$ \\
 \hline
\end{tabular*}
\caption{Allowed $LS$ $(l's')$ combinations in Eq. (\ref{eq:ls_decomposition}) for each two-body decay involved in the process $e^+ e^- \to J/\psi \, \pi^+ \, \pi^-$ and $e^+ e^- \to h_c \, \pi^+ \, \pi^-$, assuming that $Z_c$ is an axial vector meson.}
\label{Tableq1}
\end{table}
\begin{table*}[t!]
\renewcommand*{\arraystretch}{1.8}
\begin{tabular*}{\textwidth}{@{\extracolsep{\fill}}l c c c c c c}
 \hline 
 $q \ (\text{GeV})$ & $\alpha_{01} \ (\text{GeV}^2)$ & $a$ & $b \ (\text{GeV}^{-2})$ & $c$ & $d \ (\text{GeV}^{-2})$ & $\chi^2/N_{\text{dof}}$\\ 
 \hline 
 4.23 & 8.2(0.2) & 56.2(1.1) & -142.8(2.7) & -22.7(7.7) & -46.6(8.9) & 2.5\\ 
 4.26 & 6.0(0.2) & 51.8(0.9) & -132.9(2.5) & -6.9(7.4) & -46.9(8.7) & 3.1\\ 
 \hline
\end{tabular*}
\caption{Fit parameters, adjusted to reproduce the $J/\psi \, \pi^{\pm}$, $\pi^+\pi^-$ and $K^+K^-$ invariant mass distributions at $e^+e^-$ CM energies $q=4.23$ GeV and $q=4.26$ GeV. The product of $LS$ couplings $\alpha^{\gamma^* \to Z_c\pi}_{01} \cdot \alpha^{Z_c \to J/\psi\pi}_{01}$ is labeled as $\alpha_{01}$. The error indicates the uncertainties in the experimental data.}
\label{Table1}
\end{table*}

\begin{figure*}[t]
	\centering 
	\includegraphics[width=0.3\textwidth]{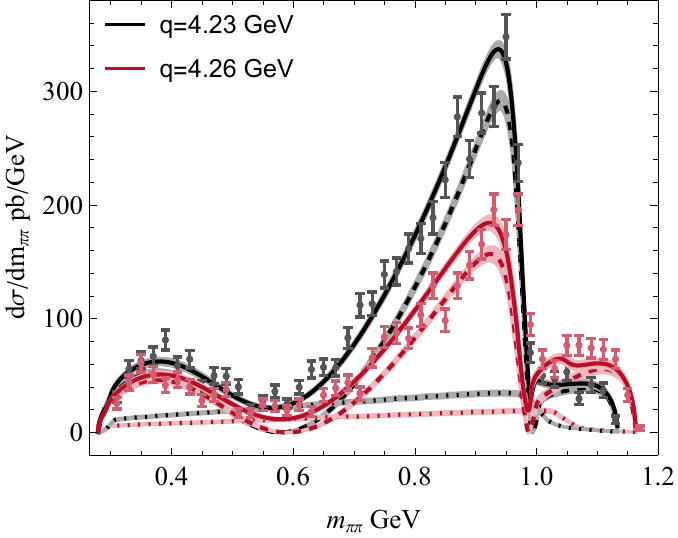} \qquad \includegraphics[width=0.3\textwidth]{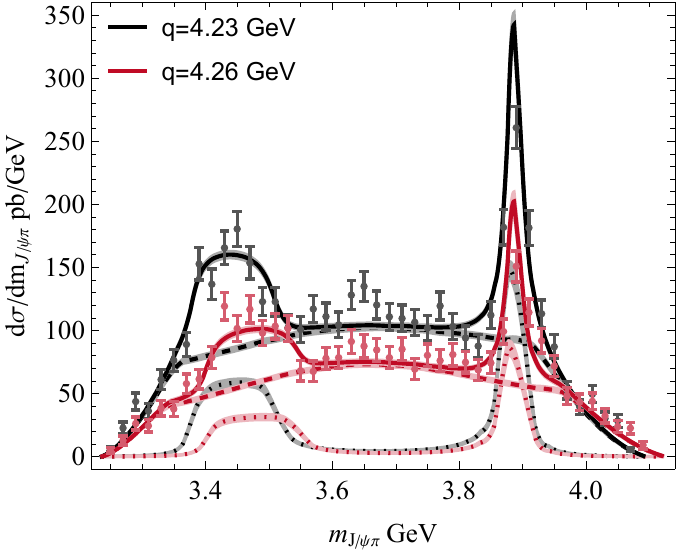} \qquad \includegraphics[width=0.3\textwidth]{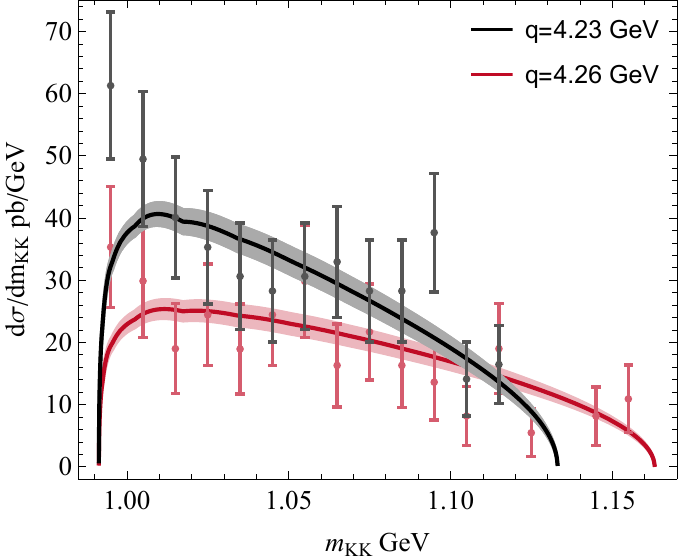}
	\caption{Invariant mass distributions of the process $e^+ e^- \to J/\psi \, \pi^+ \, \pi^- \, (K\bar{K})$ for $e^+e^-$ CM energies $q=4.23$ GeV and $q=4.26$ GeV, obtained from the minimal fit with five real parameters (see Table \ref{Table1}). For the $e^+ e^- \to J/\psi \, \pi^+ \, \pi^-$, BESIII data was taken from Ref. \cite{BESIII:2017bua}, which was normalized to the total cross-section given in Ref. \cite{BESIII:2016bnd}. Similarly, for the $e^+ e^- \to J/\psi \, K^+ K^-$, BESIII data from Ref. \cite{BESIII:2022joj} was also normalized to the respective total cross-section. The contributions from the $\pi^+\pi^-$ S-wave rescattering including $f_0(500)$ and $f_0(980)$ resonances are shown in dashed lines. The $Z^\pm_c(3900)$ resonant contributions are shown in dotted lines.}
	\label{fig_1}
\end{figure*}

\begin{figure*}[t]
	\centering 
	\includegraphics[width=0.3\textwidth]{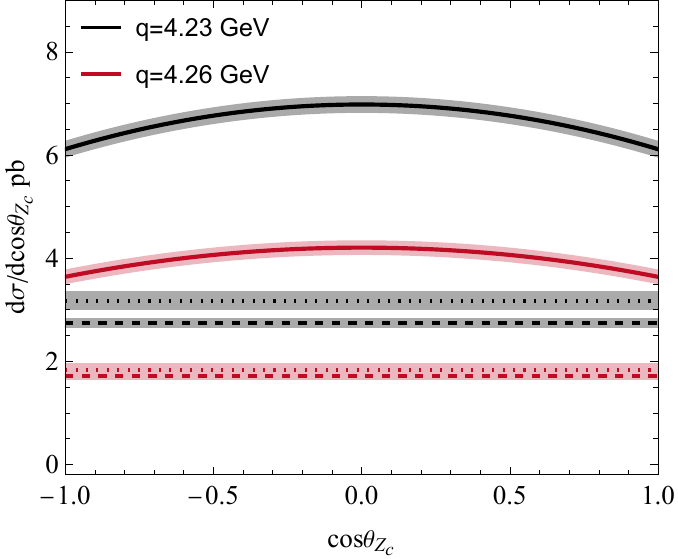} \qquad \includegraphics[width=0.3\textwidth]{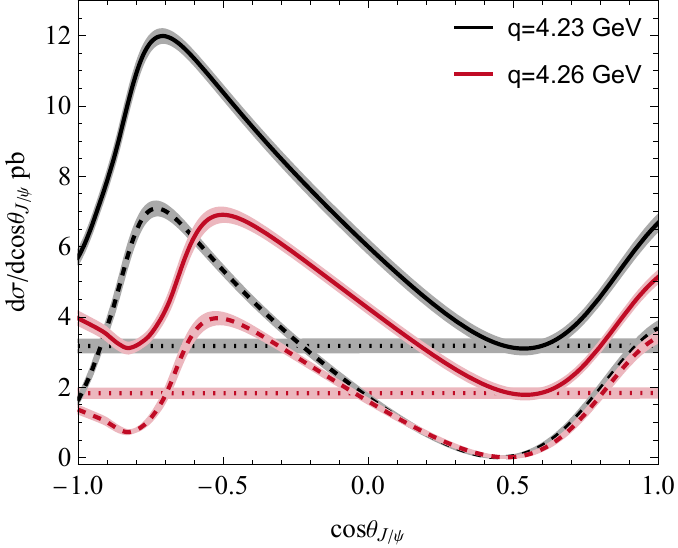}
	\caption{The polar angle distribution of $Z^\pm_c(3900)$ for the decay $\gamma^* \to Z^\pm_c \pi^\mp$ (left) and helicity angle $\theta_{J/\psi}$ distribution for the decay $Z^\pm_c \to J/\psi \pi^\pm$ (right) in the invariant mass range $m_{J/\psi \pi^\pm}\in(3.86;3.92) \ \text{GeV}$ for $e^+e^-$ CM energies $q=4.23$ GeV and $q=4.26$ GeV. The contributions from the $\pi^+\pi^-$ S-wave rescattering including $f_0(500)$ and $f_0(980)$ resonances are shown in dashed lines. The $Z^\pm_c(3900)$ resonant contributions are shown in dotted lines.}
	\label{fig_2}
\end{figure*}

\section{Numerical test}\label{5}

Using Eqs. (\ref{eq:3}) and (\ref{eq:cross_section_polarized}) for the DPD in the unpolarized $2 \to 3$ case, and a dispersive treatment for the $\pi\pi(K\bar{K})$ final state interaction (see Eqs. (\ref{eq:FSI_pipi_2}) and (\ref{eq:FSI_KK})), in this section, we present a minimal simultaneous fit to the $J/\psi \, \pi^{\pm}$, $\pi^+\pi^-$, and $K^+K^-$ invariant mass distributions of the $e^+ e^- \to J/\psi \, \pi^+ \, \pi^- \, (K^+K^-)$ process as measured by BESIII \cite{BESIII:2017bua,BESIII:2022joj}. Since we are not fitting the full Dalitz plot data sample with efficiency corrections at this stage, we have limited our analysis to the dominant S-wave contribution in all channels, which captures the main dynamics of the process and demonstrates the power of this approach. This corresponds to the lowest angular momentum in each $LS$ $(l's')$ coupling in Table \ref{Tableq1}, focusing only on the $f_0(500)/f_0(980)$ contributions. 
For the $Z_c(3900)$ contribution, we have employed the constant-width approximation, with resonance parameters fixed according to the PDG \cite{ParticleDataGroup:2022pth}. In the $S$-wave approximation, transverse and longitudinal polarizations of $Z_c$ contribute equally, i.e., there is only one unknown helicity coupling for the $\gamma^* \to Z_c\pi$ process and one for the $Z_c \to J/\psi\pi$ processes. As a result, only a few parameters are involved in the fit\footnote{For $e^+ e^-$ CM energies $q = 4.23, 4.26$ GeV, $Z_{cs}(4000)$ \cite{BESIII:2020qkh} does not appear as a peak in the $K^+K^-$ mass distribution and is therefore not taken into account.}: the product of $LS$ $(l's')$ couplings $\alpha^{\gamma^* \to Z_c\pi}_{01} \cdot \alpha^{Z_c \to J/\psi\pi}_{01}$, which characterize the $Z^{\pm}_c(3900)$ contributions (owing to Eq. (\ref{eq:crossing_relations})), and the subtraction polynomial needed for the $f_0(500)/f_0(980)$ contributions in Eqs. (\ref{eq:FSI_pipi_2}) and (\ref{eq:FSI_KK}). As shown in Fig. \ref{fig_1}, these minimal considerations are sufficient to provide a reasonable description of the invariant mass distribution data. In the future, we plan to apply this approach to the analysis of the full BESIII data samples (including the full Dalitz plot) at various $e^+e^-$ center-of-mass energies. For such an analysis, it will be necessary to include $f_2(1270)$ and D-wave contributions.

The polar angle $\theta_1$ distribution (corresponding to the polar angle of $Z^\pm_c(3900)$) for the decay $\gamma^* \to Z^\pm_c \pi^\mp$ and the helicity angle $\theta_{J/\psi}$ distribution for the decay $Z^\pm_c \to J/\psi \pi^\pm$ for distinct $e^+e^-$ CM energies in the invariant mass range $m_{J/\psi \pi^\pm}\in(3.86;3.92) \ \text{GeV}$ are shown in Fig. \ref{fig_2}. $\theta_{J/\psi}$ is defined as the angle between the momentum of $J/\psi$ in the $Z^\pm_c$ rest frame and the $Z^\pm_c$ momentum in the $e^+e^-$ rest frame. The $\pi^+\pi^-$ S-wave rescattering, accounting for the resonances $f_0(500)$ and $f_0(980)$, provides the predominant contribution to the helicity angle $\theta_{J/\psi}$ distribution.
 
For the process $e^+ e^- \to h_c \, \pi^+ \, \pi^-$, which differs from $e^+ e^- \to J/\psi \, \pi^+ \, \pi^-$ by the parity of the final charmonium state, all possible $LS$ $(l's')$ combinations are listed in Table \ref{Tableq1}. The key difference is that the lowest partial wave in the $h_c \pi^\pm$ system is the P-wave. BESIII data on $e^+ e^- \to h_c \, \pi^+ \, \pi^-$ \cite{BESIII:2013ouc} were measured at $e^+e^-$ center-of-mass energies between $3.90$ and $4.42$ GeV, indicating that the maximal physical region of the two pions does not exceed $0.9$ GeV. Therefore, the two-pion mass spectrum will be dominated by the contribution from the $f_0(500)$ resonance, which is accounted for by applying the Omnès formalism in Eq. (\ref{eq:FSI_pipi_1}). The analysis of $e^+ e^- \to h_c \, \pi \, \pi$ is currently being undertaken in cooperation with the BESIII Collaboration, with the aim of determining the spin, parity, and resonance parameters of the $Z_c(4020)$, and will be presented elsewhere.

\section{Summary and conclusions}\label{6}

In this short paper, we analyzed the $e^+ e^- \to J/\psi \, \pi \, \pi \, (K \bar{K})$ and $e^+ e^- \to h_c \, \pi \, \pi$ processes using the Dalitz-plot decomposition (DPD) approach \cite{JPAC:2019ufm}. These reactions are significant for exotic hadron searches and include the established exotic states $Z^{\pm}_c(3900)$ and $Z^{\pm}_c(4020)$. The specifics of these reactions includes the crossing symmetry of pions and the particular helicity of the intermediate virtual photon (for $e^+ e^-$ collider energies). We have shown that, for the proper treatment within the DPD framework, phase factors are necessary to address the permutation symmetry between final particles. This is crucial for implementing the $LS$ helicity coupling scheme and helps to reduce the number of unknown parameters. By incorporating a toy-model Lagrangian, we validated the factorization of the overall rotation for all decay chains and spin alignments, as well as the crossing symmetry between final states for both $1 \to 3$ and $2 \to 3$ unpolarized cases.

Furthermore, we demonstrated how to incorporate a dispersive treatment for $\pi \pi (K \bar{K})$ final state interactions in the DPD. This approach ensures consistency with the established resonances $f_0(500)$ and $f_0(980)$ and reduces the largest systematic uncertainty typically found in BESIII analyses (see e.g. \cite{BESIII:2017bua}). Using data for $e^+ e^- \to J/\psi \, \pi \, \pi \, (K \bar{K})$ from \cite{BESIII:2017bua,BESIII:2022joj}, we illustrated how a simultaneous description of invariant mass distributions can be achieved with just a few fitted parameters.
  
The results obtained are not limited to the $e^+ e^- \to J/\psi \, \pi \, \pi \, (K \bar{K})$ and $e^+ e^- \to h_c \, \pi \, \pi$ processes and can be easily applied to any $e^+ e^- \to 123$ reaction with two pions in the final state.

\section*{Acknowledgements}
We thank Y. Guo and T. Liu for helpful discussions. This work was supported by the Deutsche Forschungsgemeinschaft (DFG, German Research Foundation) within the Research Unit [Photon-photon interactions in the Standard Model and beyond, Projektnummer 458854507 - FOR 5327].

\appendix

\section{Lagrangian toy model}\label{L}

To perform verification of the DPD formalism \cite{JPAC:2019ufm}, we adopt a Lagrangian-based toy model. By assigning the corresponding Lagrangian to the interaction in each vertex, it is possible to calculate the Dalitz-plot function in two ways: directly from the Lagrangian using the polarization vector completeness relation, or through DPD, where helicity couplings are derived from the corresponding two-body decay's matrix elements using the same Lagrangian and helicity formalism. 
For the vertices involved in $\gamma^* \to J/\psi \, \pi^+ \, \pi^- \, (K^+ K^-)$ and $\gamma^* \to h_c \, \pi^+ \, \pi^-$ processes (see left column of Table \ref{Tableq1}) we utilize the following Lagrangians
\begin{equation}
\begin{split}
&\mathcal{L}_{AVP}=g_{AVP}\,\mathcal{A}_{\alpha\beta}\,\mathcal{V}^{\alpha\beta}\,\mathcal{P},\\
&\mathcal{L}_{VVS}=g_{VVS}\,\mathcal{V}_{\alpha\beta}\,\mathcal{V}^{\alpha\beta}\,\mathcal{S},\\
&\mathcal{L}_{SPP}=g_{SPP}\,\mathcal{S}\,\mathcal{P}^\dagger\,\mathcal{P},\\
&\mathcal{L}_{VVT}=g_{VVT}\,\mathcal{V}_{\eta\beta}\,\mathcal{V}^\eta_\alpha\,\mathcal{T}^{\alpha\beta},\\
&\mathcal{L}_{TPP}=g_{TPP}\,\mathcal{T}^{\alpha\beta}\,\partial_\alpha\mathcal{P}^\dagger\, \partial_\beta\mathcal{P},\\
&\mathcal{L}_{AAP}=g_{AAP}\,\tilde{G}_{\alpha\beta}\,\mathcal{A}^{\alpha\beta}\,\mathcal{P},\\
&\mathcal{L}_{AVS}=g_{AVS}\,\tilde{G}_{\alpha\beta}\,\mathcal{V}^{\alpha\beta}\,S,
\end{split}
\end{equation}
where 
\begin{equation}
\begin{split}
& \mathcal{V}_{\alpha\beta}= \partial_\alpha V_\beta - \partial_\beta V_\alpha,\\
& \mathcal{A}_{\alpha\beta}=\partial_\alpha A_\beta - \partial_\beta A_\alpha,\\
& \tilde{G}_{\alpha\beta}=\frac{1}{2}\varepsilon_{\alpha\beta\tau\eta} \left(\partial^\tau A^\eta-\partial^\eta A^\tau\right).
\end{split}
\end{equation}
Here, $\mathcal{S}$ and $\mathcal{P}$ stand for scalar and pseudoscalar fields, $V^\alpha$ 
and $A^\alpha$  denote vector and axial-vector fields, and $\mathcal{T}^{\alpha\beta}$ represents tensor field. For simplicity, we assumed that all the coupling constants $g_i=1$.

\bibliographystyle{apsrev4-1}
\bibliography{bib}

\begin{thebibliography}{50}%
\makeatletter
\providecommand \@ifxundefined [1]{%
 \@ifx{#1\undefined}
}%
\providecommand \@ifnum [1]{%
 \ifnum #1\expandafter \@firstoftwo
 \else \expandafter \@secondoftwo
 \fi
}%
\providecommand \@ifx [1]{%
 \ifx #1\expandafter \@firstoftwo
 \else \expandafter \@secondoftwo
 \fi
}%
\providecommand \natexlab [1]{#1}%
\providecommand \enquote  [1]{``#1''}%
\providecommand \bibnamefont  [1]{#1}%
\providecommand \bibfnamefont [1]{#1}%
\providecommand \citenamefont [1]{#1}%
\providecommand \href@noop [0]{\@secondoftwo}%
\providecommand \href [0]{\begingroup \@sanitize@url \@href}%
\providecommand \@href[1]{\@@startlink{#1}\@@href}%
\providecommand \@@href[1]{\endgroup#1\@@endlink}%
\providecommand \@sanitize@url [0]{\catcode `\\12\catcode `\$12\catcode `\&12\catcode `\#12\catcode `\^12\catcode `\_12\catcode `\%12\relax}%
\providecommand \@@startlink[1]{}%
\providecommand \@@endlink[0]{}%
\providecommand \url  [0]{\begingroup\@sanitize@url \@url }%
\providecommand \@url [1]{\endgroup\@href {#1}{\urlprefix }}%
\providecommand \urlprefix  [0]{URL }%
\providecommand \Eprint [0]{\href }%
\providecommand \doibase [0]{http://dx.doi.org/}%
\providecommand \selectlanguage [0]{\@gobble}%
\providecommand \bibinfo  [0]{\@secondoftwo}%
\providecommand \bibfield  [0]{\@secondoftwo}%
\providecommand \translation [1]{[#1]}%
\providecommand \BibitemOpen [0]{}%
\providecommand \bibitemStop [0]{}%
\providecommand \bibitemNoStop [0]{.\EOS\space}%
\providecommand \EOS [0]{\spacefactor3000\relax}%
\providecommand \BibitemShut  [1]{\csname bibitem#1\endcsname}%
\let\auto@bib@innerbib\@empty
%</preamble>
\bibitem [{\citenamefont {Rarita}\ and\ \citenamefont {Schwinger}(1941)}]{Rarita:1941mf}%
  \BibitemOpen
  \bibfield  {author} {\bibinfo {author} {\bibfnamefont {W.}~\bibnamefont {Rarita}}\ and\ \bibinfo {author} {\bibfnamefont {J.}~\bibnamefont {Schwinger}},\ }\href {\doibase 10.1103/PhysRev.60.61} {\bibfield  {journal} {\bibinfo  {journal} {Phys. Rev.}\ }\textbf {\bibinfo {volume} {60}},\ \bibinfo {pages} {61} (\bibinfo {year} {1941})}\BibitemShut {NoStop}%
\bibitem [{\citenamefont {Zemach}(1964)}]{Zemach:1963bc}%
  \BibitemOpen
  \bibfield  {author} {\bibinfo {author} {\bibfnamefont {C.}~\bibnamefont {Zemach}},\ }\href {\doibase 10.1103/PhysRev.133.B1201} {\bibfield  {journal} {\bibinfo  {journal} {Phys. Rev.}\ }\textbf {\bibinfo {volume} {133}},\ \bibinfo {pages} {B1201} (\bibinfo {year} {1964})}\BibitemShut {NoStop}%
\bibitem [{\citenamefont {Zemach}(1965)}]{Zemach:1965ycj}%
  \BibitemOpen
  \bibfield  {author} {\bibinfo {author} {\bibfnamefont {C.}~\bibnamefont {Zemach}},\ }\href {\doibase 10.1103/PhysRev.140.B97} {\bibfield  {journal} {\bibinfo  {journal} {Phys. Rev.}\ }\textbf {\bibinfo {volume} {140}},\ \bibinfo {pages} {B97} (\bibinfo {year} {1965})}\BibitemShut {NoStop}%
\bibitem [{\citenamefont {Jacob}\ and\ \citenamefont {Wick}(1959)}]{Jacob:1959at}%
  \BibitemOpen
  \bibfield  {author} {\bibinfo {author} {\bibfnamefont {M.}~\bibnamefont {Jacob}}\ and\ \bibinfo {author} {\bibfnamefont {G.~C.}\ \bibnamefont {Wick}},\ }\href {\doibase 10.1006/aphy.2000.6022} {\bibfield  {journal} {\bibinfo  {journal} {Annals Phys.}\ }\textbf {\bibinfo {volume} {7}},\ \bibinfo {pages} {404} (\bibinfo {year} {1959})}\BibitemShut {NoStop}%
\bibitem [{\citenamefont {Chung}\ and\ \citenamefont {Friedrich}(2008)}]{Chung:2007nn}%
  \BibitemOpen
  \bibfield  {author} {\bibinfo {author} {\bibfnamefont {S.-U.}\ \bibnamefont {Chung}}\ and\ \bibinfo {author} {\bibfnamefont {J.}~\bibnamefont {Friedrich}},\ }\href {\doibase 10.1103/PhysRevD.78.074027} {\bibfield  {journal} {\bibinfo  {journal} {Phys. Rev. D}\ }\textbf {\bibinfo {volume} {78}},\ \bibinfo {pages} {074027} (\bibinfo {year} {2008})},\ \Eprint {http://arxiv.org/abs/0711.3143} {arXiv:0711.3143 [hep-ph]} \BibitemShut {NoStop}%
\bibitem [{\citenamefont {Chen}\ and\ \citenamefont {Ping}(2017)}]{Chen:2017gtx}%
  \BibitemOpen
  \bibfield  {author} {\bibinfo {author} {\bibfnamefont {H.}~\bibnamefont {Chen}}\ and\ \bibinfo {author} {\bibfnamefont {R.-G.}\ \bibnamefont {Ping}},\ }\href {\doibase 10.1103/PhysRevD.95.076010} {\bibfield  {journal} {\bibinfo  {journal} {Phys. Rev. D}\ }\textbf {\bibinfo {volume} {95}},\ \bibinfo {pages} {076010} (\bibinfo {year} {2017})},\ \Eprint {http://arxiv.org/abs/1704.05184} {arXiv:1704.05184 [hep-ph]} \BibitemShut {NoStop}%
\bibitem [{\citenamefont {Mikhasenko}\ \emph {et~al.}(2020)\citenamefont {Mikhasenko} \emph {et~al.}}]{JPAC:2019ufm}%
  \BibitemOpen
  \bibfield  {author} {\bibinfo {author} {\bibfnamefont {M.}~\bibnamefont {Mikhasenko}} \emph {et~al.} (\bibinfo {collaboration} {JPAC}),\ }\href {\doibase 10.1103/PhysRevD.101.034033} {\bibfield  {journal} {\bibinfo  {journal} {Phys. Rev. D}\ }\textbf {\bibinfo {volume} {101}},\ \bibinfo {pages} {034033} (\bibinfo {year} {2020})},\ \Eprint {http://arxiv.org/abs/1910.04566} {arXiv:1910.04566 [hep-ph]} \BibitemShut {NoStop}%
\bibitem [{\citenamefont {Habermann}\ and\ \citenamefont {Mikhasenko}(2024)}]{Habermann:2024sxs}%
  \BibitemOpen
  \bibfield  {author} {\bibinfo {author} {\bibfnamefont {K.}~\bibnamefont {Habermann}}\ and\ \bibinfo {author} {\bibfnamefont {M.}~\bibnamefont {Mikhasenko}},\ }\href@noop {} {\  (\bibinfo {year} {2024})},\ \Eprint {http://arxiv.org/abs/2409.06913} {arXiv:2409.06913 [hep-ph]} \BibitemShut {NoStop}%
\bibitem [{\citenamefont {Brambilla}\ \emph {et~al.}(2020)\citenamefont {Brambilla}, \citenamefont {Eidelman}, \citenamefont {Hanhart}, \citenamefont {Nefediev}, \citenamefont {Shen}, \citenamefont {Thomas}, \citenamefont {Vairo},\ and\ \citenamefont {Yuan}}]{Brambilla:2019esw}%
  \BibitemOpen
  \bibfield  {author} {\bibinfo {author} {\bibfnamefont {N.}~\bibnamefont {Brambilla}}, \bibinfo {author} {\bibfnamefont {S.}~\bibnamefont {Eidelman}}, \bibinfo {author} {\bibfnamefont {C.}~\bibnamefont {Hanhart}}, \bibinfo {author} {\bibfnamefont {A.}~\bibnamefont {Nefediev}}, \bibinfo {author} {\bibfnamefont {C.-P.}\ \bibnamefont {Shen}}, \bibinfo {author} {\bibfnamefont {C.~E.}\ \bibnamefont {Thomas}}, \bibinfo {author} {\bibfnamefont {A.}~\bibnamefont {Vairo}}, \ and\ \bibinfo {author} {\bibfnamefont {C.-Z.}\ \bibnamefont {Yuan}},\ }\href {\doibase 10.1016/j.physrep.2020.05.001} {\bibfield  {journal} {\bibinfo  {journal} {Phys. Rept.}\ }\textbf {\bibinfo {volume} {873}},\ \bibinfo {pages} {1} (\bibinfo {year} {2020})},\ \Eprint {http://arxiv.org/abs/1907.07583} {arXiv:1907.07583 [hep-ex]} \BibitemShut {NoStop}%
\bibitem [{\citenamefont {Chen}\ \emph {et~al.}(2016)\citenamefont {Chen}, \citenamefont {Chen}, \citenamefont {Liu},\ and\ \citenamefont {Zhu}}]{Chen:2016qju}%
  \BibitemOpen
  \bibfield  {author} {\bibinfo {author} {\bibfnamefont {H.-X.}\ \bibnamefont {Chen}}, \bibinfo {author} {\bibfnamefont {W.}~\bibnamefont {Chen}}, \bibinfo {author} {\bibfnamefont {X.}~\bibnamefont {Liu}}, \ and\ \bibinfo {author} {\bibfnamefont {S.-L.}\ \bibnamefont {Zhu}},\ }\href {\doibase 10.1016/j.physrep.2016.05.004} {\bibfield  {journal} {\bibinfo  {journal} {Phys. Rept.}\ }\textbf {\bibinfo {volume} {639}},\ \bibinfo {pages} {1} (\bibinfo {year} {2016})},\ \Eprint {http://arxiv.org/abs/1601.02092} {arXiv:1601.02092 [hep-ph]} \BibitemShut {NoStop}%
\bibitem [{\citenamefont {Chen}\ \emph {et~al.}(2023)\citenamefont {Chen}, \citenamefont {Chen}, \citenamefont {Liu}, \citenamefont {Liu},\ and\ \citenamefont {Zhu}}]{Chen:2022asf}%
  \BibitemOpen
  \bibfield  {author} {\bibinfo {author} {\bibfnamefont {H.-X.}\ \bibnamefont {Chen}}, \bibinfo {author} {\bibfnamefont {W.}~\bibnamefont {Chen}}, \bibinfo {author} {\bibfnamefont {X.}~\bibnamefont {Liu}}, \bibinfo {author} {\bibfnamefont {Y.-R.}\ \bibnamefont {Liu}}, \ and\ \bibinfo {author} {\bibfnamefont {S.-L.}\ \bibnamefont {Zhu}},\ }\href {\doibase 10.1088/1361-6633/aca3b6} {\bibfield  {journal} {\bibinfo  {journal} {Rept. Prog. Phys.}\ }\textbf {\bibinfo {volume} {86}},\ \bibinfo {pages} {026201} (\bibinfo {year} {2023})},\ \Eprint {http://arxiv.org/abs/2204.02649} {arXiv:2204.02649 [hep-ph]} \BibitemShut {NoStop}%
\bibitem [{\citenamefont {Olsen}\ \emph {et~al.}(2018)\citenamefont {Olsen}, \citenamefont {Skwarnicki},\ and\ \citenamefont {Zieminska}}]{Olsen:2017bmm}%
  \BibitemOpen
  \bibfield  {author} {\bibinfo {author} {\bibfnamefont {S.~L.}\ \bibnamefont {Olsen}}, \bibinfo {author} {\bibfnamefont {T.}~\bibnamefont {Skwarnicki}}, \ and\ \bibinfo {author} {\bibfnamefont {D.}~\bibnamefont {Zieminska}},\ }\href {\doibase 10.1103/RevModPhys.90.015003} {\bibfield  {journal} {\bibinfo  {journal} {Rev. Mod. Phys.}\ }\textbf {\bibinfo {volume} {90}},\ \bibinfo {pages} {015003} (\bibinfo {year} {2018})},\ \Eprint {http://arxiv.org/abs/1708.04012} {arXiv:1708.04012 [hep-ph]} \BibitemShut {NoStop}%
\bibitem [{\citenamefont {Karliner}\ \emph {et~al.}(2018)\citenamefont {Karliner}, \citenamefont {Rosner},\ and\ \citenamefont {Skwarnicki}}]{Karliner:2017qhf}%
  \BibitemOpen
  \bibfield  {author} {\bibinfo {author} {\bibfnamefont {M.}~\bibnamefont {Karliner}}, \bibinfo {author} {\bibfnamefont {J.~L.}\ \bibnamefont {Rosner}}, \ and\ \bibinfo {author} {\bibfnamefont {T.}~\bibnamefont {Skwarnicki}},\ }\href {\doibase 10.1146/annurev-nucl-101917-020902} {\bibfield  {journal} {\bibinfo  {journal} {Ann. Rev. Nucl. Part. Sci.}\ }\textbf {\bibinfo {volume} {68}},\ \bibinfo {pages} {17} (\bibinfo {year} {2018})},\ \Eprint {http://arxiv.org/abs/1711.10626} {arXiv:1711.10626 [hep-ph]} \BibitemShut {NoStop}%
\bibitem [{\citenamefont {Guo}\ \emph {et~al.}(2018)\citenamefont {Guo}, \citenamefont {Hanhart}, \citenamefont {Mei\ss{}ner}, \citenamefont {Wang}, \citenamefont {Zhao},\ and\ \citenamefont {Zou}}]{Guo:2017jvc}%
  \BibitemOpen
  \bibfield  {author} {\bibinfo {author} {\bibfnamefont {F.-K.}\ \bibnamefont {Guo}}, \bibinfo {author} {\bibfnamefont {C.}~\bibnamefont {Hanhart}}, \bibinfo {author} {\bibfnamefont {U.-G.}\ \bibnamefont {Mei\ss{}ner}}, \bibinfo {author} {\bibfnamefont {Q.}~\bibnamefont {Wang}}, \bibinfo {author} {\bibfnamefont {Q.}~\bibnamefont {Zhao}}, \ and\ \bibinfo {author} {\bibfnamefont {B.-S.}\ \bibnamefont {Zou}},\ }\href {\doibase 10.1103/RevModPhys.90.015004} {\bibfield  {journal} {\bibinfo  {journal} {Rev. Mod. Phys.}\ }\textbf {\bibinfo {volume} {90}},\ \bibinfo {pages} {015004} (\bibinfo {year} {2018})},\ \bibinfo {note} {[Erratum: Rev.Mod.Phys. 94, 029901 (2022)]},\ \Eprint {http://arxiv.org/abs/1705.00141} {arXiv:1705.00141 [hep-ph]} \BibitemShut {NoStop}%
\bibitem [{\citenamefont {Liu}\ \emph {et~al.}(2019)\citenamefont {Liu}, \citenamefont {Chen}, \citenamefont {Chen}, \citenamefont {Liu},\ and\ \citenamefont {Zhu}}]{Liu:2019zoy}%
  \BibitemOpen
  \bibfield  {author} {\bibinfo {author} {\bibfnamefont {Y.-R.}\ \bibnamefont {Liu}}, \bibinfo {author} {\bibfnamefont {H.-X.}\ \bibnamefont {Chen}}, \bibinfo {author} {\bibfnamefont {W.}~\bibnamefont {Chen}}, \bibinfo {author} {\bibfnamefont {X.}~\bibnamefont {Liu}}, \ and\ \bibinfo {author} {\bibfnamefont {S.-L.}\ \bibnamefont {Zhu}},\ }\href {\doibase 10.1016/j.ppnp.2019.04.003} {\bibfield  {journal} {\bibinfo  {journal} {Prog. Part. Nucl. Phys.}\ }\textbf {\bibinfo {volume} {107}},\ \bibinfo {pages} {237} (\bibinfo {year} {2019})},\ \Eprint {http://arxiv.org/abs/1903.11976} {arXiv:1903.11976 [hep-ph]} \BibitemShut {NoStop}%
\bibitem [{\citenamefont {Garcia-Martin}\ \emph {et~al.}(2011{\natexlab{a}})\citenamefont {Garcia-Martin}, \citenamefont {Kaminski}, \citenamefont {Pelaez}, \citenamefont {Ruiz~de Elvira},\ and\ \citenamefont {Yndurain}}]{Garcia-Martin:2011iqs}%
  \BibitemOpen
  \bibfield  {author} {\bibinfo {author} {\bibfnamefont {R.}~\bibnamefont {Garcia-Martin}}, \bibinfo {author} {\bibfnamefont {R.}~\bibnamefont {Kaminski}}, \bibinfo {author} {\bibfnamefont {J.~R.}\ \bibnamefont {Pelaez}}, \bibinfo {author} {\bibfnamefont {J.}~\bibnamefont {Ruiz~de Elvira}}, \ and\ \bibinfo {author} {\bibfnamefont {F.~J.}\ \bibnamefont {Yndurain}},\ }\href {\doibase 10.1103/PhysRevD.83.074004} {\bibfield  {journal} {\bibinfo  {journal} {Phys. Rev. D}\ }\textbf {\bibinfo {volume} {83}},\ \bibinfo {pages} {074004} (\bibinfo {year} {2011}{\natexlab{a}})},\ \Eprint {http://arxiv.org/abs/1102.2183} {arXiv:1102.2183 [hep-ph]} \BibitemShut {NoStop}%
\bibitem [{\citenamefont {Buettiker}\ \emph {et~al.}(2004)\citenamefont {Buettiker}, \citenamefont {Descotes-Genon},\ and\ \citenamefont {Moussallam}}]{Buettiker:2003pp}%
  \BibitemOpen
  \bibfield  {author} {\bibinfo {author} {\bibfnamefont {P.}~\bibnamefont {Buettiker}}, \bibinfo {author} {\bibfnamefont {S.}~\bibnamefont {Descotes-Genon}}, \ and\ \bibinfo {author} {\bibfnamefont {B.}~\bibnamefont {Moussallam}},\ }\href {\doibase 10.1140/epjc/s2004-01591-1} {\bibfield  {journal} {\bibinfo  {journal} {Eur. Phys. J. C}\ }\textbf {\bibinfo {volume} {33}},\ \bibinfo {pages} {409} (\bibinfo {year} {2004})},\ \Eprint {http://arxiv.org/abs/hep-ph/0310283} {arXiv:hep-ph/0310283} \BibitemShut {NoStop}%
\bibitem [{\citenamefont {Pelaez}\ and\ \citenamefont {Rodas}(2018)}]{Pelaez:2018qny}%
  \BibitemOpen
  \bibfield  {author} {\bibinfo {author} {\bibfnamefont {J.~R.}\ \bibnamefont {Pelaez}}\ and\ \bibinfo {author} {\bibfnamefont {A.}~\bibnamefont {Rodas}},\ }\href {\doibase 10.1140/epjc/s10052-018-6296-9} {\bibfield  {journal} {\bibinfo  {journal} {Eur. Phys. J. C}\ }\textbf {\bibinfo {volume} {78}},\ \bibinfo {pages} {897} (\bibinfo {year} {2018})},\ \Eprint {http://arxiv.org/abs/1807.04543} {arXiv:1807.04543 [hep-ph]} \BibitemShut {NoStop}%
\bibitem [{\citenamefont {Pilloni}\ \emph {et~al.}(2017)\citenamefont {Pilloni}, \citenamefont {Fernandez-Ramirez}, \citenamefont {Jackura}, \citenamefont {Mathieu}, \citenamefont {Mikhasenko}, \citenamefont {Nys},\ and\ \citenamefont {Szczepaniak}}]{Pilloni:2016obd}%
  \BibitemOpen
  \bibfield  {author} {\bibinfo {author} {\bibfnamefont {A.}~\bibnamefont {Pilloni}}, \bibinfo {author} {\bibfnamefont {C.}~\bibnamefont {Fernandez-Ramirez}}, \bibinfo {author} {\bibfnamefont {A.}~\bibnamefont {Jackura}}, \bibinfo {author} {\bibfnamefont {V.}~\bibnamefont {Mathieu}}, \bibinfo {author} {\bibfnamefont {M.}~\bibnamefont {Mikhasenko}}, \bibinfo {author} {\bibfnamefont {J.}~\bibnamefont {Nys}}, \ and\ \bibinfo {author} {\bibfnamefont {A.~P.}\ \bibnamefont {Szczepaniak}} (\bibinfo {collaboration} {JPAC}),\ }\href {\doibase 10.1016/j.physletb.2017.06.030} {\bibfield  {journal} {\bibinfo  {journal} {Phys. Lett. B}\ }\textbf {\bibinfo {volume} {772}},\ \bibinfo {pages} {200} (\bibinfo {year} {2017})},\ \Eprint {http://arxiv.org/abs/1612.06490} {arXiv:1612.06490 [hep-ph]} \BibitemShut {NoStop}%
\bibitem [{\citenamefont {Ablikim}\ \emph {et~al.}(2017{\natexlab{a}})\citenamefont {Ablikim} \emph {et~al.}}]{BESIII:2017bua}%
  \BibitemOpen
  \bibfield  {author} {\bibinfo {author} {\bibfnamefont {M.}~\bibnamefont {Ablikim}} \emph {et~al.} (\bibinfo {collaboration} {BESIII}),\ }\href {\doibase 10.1103/PhysRevLett.119.072001} {\bibfield  {journal} {\bibinfo  {journal} {Phys. Rev. Lett.}\ }\textbf {\bibinfo {volume} {119}},\ \bibinfo {pages} {072001} (\bibinfo {year} {2017}{\natexlab{a}})},\ \Eprint {http://arxiv.org/abs/1706.04100} {arXiv:1706.04100 [hep-ex]} \BibitemShut {NoStop}%
\bibitem [{\citenamefont {Ablikim}\ \emph {et~al.}(2020)\citenamefont {Ablikim} \emph {et~al.}}]{BESIII:2020oph}%
  \BibitemOpen
  \bibfield  {author} {\bibinfo {author} {\bibfnamefont {M.}~\bibnamefont {Ablikim}} \emph {et~al.} (\bibinfo {collaboration} {BESIII}),\ }\href {\doibase 10.1103/PhysRevD.102.012009} {\bibfield  {journal} {\bibinfo  {journal} {Phys. Rev. D}\ }\textbf {\bibinfo {volume} {102}},\ \bibinfo {pages} {012009} (\bibinfo {year} {2020})},\ \Eprint {http://arxiv.org/abs/2004.13788} {arXiv:2004.13788 [hep-ex]} \BibitemShut {NoStop}%
\bibitem [{\citenamefont {Danilkin}\ \emph {et~al.}(2020)\citenamefont {Danilkin}, \citenamefont {Molnar},\ and\ \citenamefont {Vanderhaeghen}}]{Danilkin:2020kce}%
  \BibitemOpen
  \bibfield  {author} {\bibinfo {author} {\bibfnamefont {I.}~\bibnamefont {Danilkin}}, \bibinfo {author} {\bibfnamefont {D.~A.~S.}\ \bibnamefont {Molnar}}, \ and\ \bibinfo {author} {\bibfnamefont {M.}~\bibnamefont {Vanderhaeghen}},\ }\href {\doibase 10.1103/PhysRevD.102.016019} {\bibfield  {journal} {\bibinfo  {journal} {Phys. Rev. D}\ }\textbf {\bibinfo {volume} {102}},\ \bibinfo {pages} {016019} (\bibinfo {year} {2020})},\ \Eprint {http://arxiv.org/abs/2004.13499} {arXiv:2004.13499 [hep-ph]} \BibitemShut {NoStop}%
\bibitem [{\citenamefont {Danilkin}\ \emph {et~al.}(2021)\citenamefont {Danilkin}, \citenamefont {Deineka},\ and\ \citenamefont {Vanderhaeghen}}]{Danilkin:2020pak}%
  \BibitemOpen
  \bibfield  {author} {\bibinfo {author} {\bibfnamefont {I.}~\bibnamefont {Danilkin}}, \bibinfo {author} {\bibfnamefont {O.}~\bibnamefont {Deineka}}, \ and\ \bibinfo {author} {\bibfnamefont {M.}~\bibnamefont {Vanderhaeghen}},\ }\href {\doibase 10.1103/PhysRevD.103.114023} {\bibfield  {journal} {\bibinfo  {journal} {Phys. Rev. D}\ }\textbf {\bibinfo {volume} {103}},\ \bibinfo {pages} {114023} (\bibinfo {year} {2021})},\ \Eprint {http://arxiv.org/abs/2012.11636} {arXiv:2012.11636 [hep-ph]} \BibitemShut {NoStop}%
\bibitem [{\citenamefont {Ablikim}\ \emph {et~al.}(2022)\citenamefont {Ablikim} \emph {et~al.}}]{BESIII:2022joj}%
  \BibitemOpen
  \bibfield  {author} {\bibinfo {author} {\bibfnamefont {M.}~\bibnamefont {Ablikim}} \emph {et~al.} (\bibinfo {collaboration} {BESIII}),\ }\href {\doibase 10.1088/1674-1137/ac945c} {\bibfield  {journal} {\bibinfo  {journal} {Chin. Phys. C}\ }\textbf {\bibinfo {volume} {46}},\ \bibinfo {pages} {111002} (\bibinfo {year} {2022})},\ \Eprint {http://arxiv.org/abs/2204.07800} {arXiv:2204.07800 [hep-ex]} \BibitemShut {NoStop}%
\bibitem [{\citenamefont {Chung}(1998)}]{Chung:1997jn}%
  \BibitemOpen
  \bibfield  {author} {\bibinfo {author} {\bibfnamefont {S.~U.}\ \bibnamefont {Chung}},\ }\href {\doibase 10.1103/PhysRevD.57.431} {\bibfield  {journal} {\bibinfo  {journal} {Phys. Rev. D}\ }\textbf {\bibinfo {volume} {57}},\ \bibinfo {pages} {431} (\bibinfo {year} {1998})}\BibitemShut {NoStop}%
\bibitem [{\citenamefont {Zou}\ and\ \citenamefont {Bugg}(2003)}]{Zou:2002ar}%
  \BibitemOpen
  \bibfield  {author} {\bibinfo {author} {\bibfnamefont {B.~S.}\ \bibnamefont {Zou}}\ and\ \bibinfo {author} {\bibfnamefont {D.~V.}\ \bibnamefont {Bugg}},\ }\href {\doibase 10.1140/epja/i2002-10135-4} {\bibfield  {journal} {\bibinfo  {journal} {Eur. Phys. J. A}\ }\textbf {\bibinfo {volume} {16}},\ \bibinfo {pages} {537} (\bibinfo {year} {2003})},\ \Eprint {http://arxiv.org/abs/hep-ph/0211457} {arXiv:hep-ph/0211457} \BibitemShut {NoStop}%
\bibitem [{\citenamefont {Von~Hippel}\ and\ \citenamefont {Quigg}(1972)}]{VonHippel:1972fg}%
  \BibitemOpen
  \bibfield  {author} {\bibinfo {author} {\bibfnamefont {F.}~\bibnamefont {Von~Hippel}}\ and\ \bibinfo {author} {\bibfnamefont {C.}~\bibnamefont {Quigg}},\ }\href {\doibase 10.1103/PhysRevD.5.624} {\bibfield  {journal} {\bibinfo  {journal} {Phys. Rev. D}\ }\textbf {\bibinfo {volume} {5}},\ \bibinfo {pages} {624} (\bibinfo {year} {1972})}\BibitemShut {NoStop}%
\bibitem [{\citenamefont {Liu}\ \emph {et~al.}(2013)\citenamefont {Liu} \emph {et~al.}}]{Belle:2013yex}%
  \BibitemOpen
  \bibfield  {author} {\bibinfo {author} {\bibfnamefont {Z.~Q.}\ \bibnamefont {Liu}} \emph {et~al.} (\bibinfo {collaboration} {Belle}),\ }\href {\doibase 10.1103/PhysRevLett.110.252002} {\bibfield  {journal} {\bibinfo  {journal} {Phys. Rev. Lett.}\ }\textbf {\bibinfo {volume} {110}},\ \bibinfo {pages} {252002} (\bibinfo {year} {2013})},\ \bibinfo {note} {[Erratum: Phys.Rev.Lett. 111, 019901 (2013)]},\ \Eprint {http://arxiv.org/abs/1304.0121} {arXiv:1304.0121 [hep-ex]} \BibitemShut {NoStop}%
\bibitem [{\citenamefont {Ablikim}\ \emph {et~al.}(2014)\citenamefont {Ablikim} \emph {et~al.}}]{BESIII:2013mhi}%
  \BibitemOpen
  \bibfield  {author} {\bibinfo {author} {\bibfnamefont {M.}~\bibnamefont {Ablikim}} \emph {et~al.} (\bibinfo {collaboration} {BESIII}),\ }\href {\doibase 10.1103/PhysRevLett.112.132001} {\bibfield  {journal} {\bibinfo  {journal} {Phys. Rev. Lett.}\ }\textbf {\bibinfo {volume} {112}},\ \bibinfo {pages} {132001} (\bibinfo {year} {2014})},\ \Eprint {http://arxiv.org/abs/1308.2760} {arXiv:1308.2760 [hep-ex]} \BibitemShut {NoStop}%
\bibitem [{\citenamefont {Ablikim}\ \emph {et~al.}(2013)\citenamefont {Ablikim} \emph {et~al.}}]{BESIII:2013ouc}%
  \BibitemOpen
  \bibfield  {author} {\bibinfo {author} {\bibfnamefont {M.}~\bibnamefont {Ablikim}} \emph {et~al.} (\bibinfo {collaboration} {BESIII}),\ }\href {\doibase 10.1103/PhysRevLett.111.242001} {\bibfield  {journal} {\bibinfo  {journal} {Phys. Rev. Lett.}\ }\textbf {\bibinfo {volume} {111}},\ \bibinfo {pages} {242001} (\bibinfo {year} {2013})},\ \Eprint {http://arxiv.org/abs/1309.1896} {arXiv:1309.1896 [hep-ex]} \BibitemShut {NoStop}%
\bibitem [{\citenamefont {Chung}(1971)}]{Chung:1971ri}%
  \BibitemOpen
  \bibfield  {author} {\bibinfo {author} {\bibfnamefont {S.~U.}\ \bibnamefont {Chung}},\ }\href {\doibase 10.5170/CERN-1971-008} {\bibfield  {journal} {\bibinfo  {journal} {CERN-71-08}\ } (\bibinfo {year} {1971}),\ 10.5170/CERN-1971-008}\BibitemShut {NoStop}%
\bibitem [{\citenamefont {Albaladejo}\ \emph {et~al.}(2022)\citenamefont {Albaladejo} \emph {et~al.}}]{JPAC:2021rxu}%
  \BibitemOpen
  \bibfield  {author} {\bibinfo {author} {\bibfnamefont {M.}~\bibnamefont {Albaladejo}} \emph {et~al.} (\bibinfo {collaboration} {JPAC}),\ }\href {\doibase 10.1016/j.ppnp.2022.103981} {\bibfield  {journal} {\bibinfo  {journal} {Prog. Part. Nucl. Phys.}\ }\textbf {\bibinfo {volume} {127}},\ \bibinfo {pages} {103981} (\bibinfo {year} {2022})},\ \Eprint {http://arxiv.org/abs/2112.13436} {arXiv:2112.13436 [hep-ph]} \BibitemShut {NoStop}%
\bibitem [{\citenamefont {Khuri}\ and\ \citenamefont {Treiman}(1960)}]{Khuri:1960zz}%
  \BibitemOpen
  \bibfield  {author} {\bibinfo {author} {\bibfnamefont {N.~N.}\ \bibnamefont {Khuri}}\ and\ \bibinfo {author} {\bibfnamefont {S.~B.}\ \bibnamefont {Treiman}},\ }\href {\doibase 10.1103/PhysRev.119.1115} {\bibfield  {journal} {\bibinfo  {journal} {Phys. Rev.}\ }\textbf {\bibinfo {volume} {119}},\ \bibinfo {pages} {1115} (\bibinfo {year} {1960})}\BibitemShut {NoStop}%
\bibitem [{\citenamefont {Albaladejo}\ and\ \citenamefont {Moussallam}(2017)}]{Albaladejo:2017hhj}%
  \BibitemOpen
  \bibfield  {author} {\bibinfo {author} {\bibfnamefont {M.}~\bibnamefont {Albaladejo}}\ and\ \bibinfo {author} {\bibfnamefont {B.}~\bibnamefont {Moussallam}},\ }\href {\doibase 10.1140/epjc/s10052-017-5052-x} {\bibfield  {journal} {\bibinfo  {journal} {Eur. Phys. J. C}\ }\textbf {\bibinfo {volume} {77}},\ \bibinfo {pages} {508} (\bibinfo {year} {2017})},\ \Eprint {http://arxiv.org/abs/1702.04931} {arXiv:1702.04931 [hep-ph]} \BibitemShut {NoStop}%
\bibitem [{\citenamefont {Albaladejo}\ \emph {et~al.}(2020)\citenamefont {Albaladejo}, \citenamefont {Winney}, \citenamefont {Danilkin}, \citenamefont {Fern\'andez-Ram\'\i{}rez}, \citenamefont {Mathieu}, \citenamefont {Mikhasenko}, \citenamefont {Pilloni}, \citenamefont {Silva-Castro},\ and\ \citenamefont {Szczepaniak}}]{Albaladejo:2019huw}%
  \BibitemOpen
  \bibfield  {author} {\bibinfo {author} {\bibfnamefont {M.}~\bibnamefont {Albaladejo}}, \bibinfo {author} {\bibfnamefont {D.}~\bibnamefont {Winney}}, \bibinfo {author} {\bibfnamefont {I.~V.}\ \bibnamefont {Danilkin}}, \bibinfo {author} {\bibfnamefont {C.}~\bibnamefont {Fern\'andez-Ram\'\i{}rez}}, \bibinfo {author} {\bibfnamefont {V.}~\bibnamefont {Mathieu}}, \bibinfo {author} {\bibfnamefont {M.}~\bibnamefont {Mikhasenko}}, \bibinfo {author} {\bibfnamefont {A.}~\bibnamefont {Pilloni}}, \bibinfo {author} {\bibfnamefont {J.~A.}\ \bibnamefont {Silva-Castro}}, \ and\ \bibinfo {author} {\bibfnamefont {A.~P.}\ \bibnamefont {Szczepaniak}} (\bibinfo {collaboration} {JPAC}),\ }\href {\doibase 10.1103/PhysRevD.101.054018} {\bibfield  {journal} {\bibinfo  {journal} {Phys. Rev. D}\ }\textbf {\bibinfo {volume} {101}},\ \bibinfo {pages} {054018} (\bibinfo {year} {2020})},\ \Eprint {http://arxiv.org/abs/1910.03107} {arXiv:1910.03107 [hep-ph]} \BibitemShut {NoStop}%
\bibitem [{\citenamefont {Stamen}\ \emph {et~al.}(2023)\citenamefont {Stamen}, \citenamefont {Isken}, \citenamefont {Kubis}, \citenamefont {Mikhasenko},\ and\ \citenamefont {Niehus}}]{Stamen:2022eda}%
  \BibitemOpen
  \bibfield  {author} {\bibinfo {author} {\bibfnamefont {D.}~\bibnamefont {Stamen}}, \bibinfo {author} {\bibfnamefont {T.}~\bibnamefont {Isken}}, \bibinfo {author} {\bibfnamefont {B.}~\bibnamefont {Kubis}}, \bibinfo {author} {\bibfnamefont {M.}~\bibnamefont {Mikhasenko}}, \ and\ \bibinfo {author} {\bibfnamefont {M.}~\bibnamefont {Niehus}},\ }\href {\doibase 10.1140/epjc/s10052-023-11665-x} {\bibfield  {journal} {\bibinfo  {journal} {Eur. Phys. J. C}\ }\textbf {\bibinfo {volume} {83}},\ \bibinfo {pages} {510} (\bibinfo {year} {2023})},\ \bibinfo {note} {[Erratum: Eur.Phys.J.C 83, 586 (2023)]},\ \Eprint {http://arxiv.org/abs/2212.11767} {arXiv:2212.11767 [hep-ph]} \BibitemShut {NoStop}%
\bibitem [{\citenamefont {Danilkin}\ \emph {et~al.}(2023)\citenamefont {Danilkin}, \citenamefont {Biloshytskyi}, \citenamefont {Ren},\ and\ \citenamefont {Vanderhaeghen}}]{Danilkin:2022cnj}%
  \BibitemOpen
  \bibfield  {author} {\bibinfo {author} {\bibfnamefont {I.}~\bibnamefont {Danilkin}}, \bibinfo {author} {\bibfnamefont {V.}~\bibnamefont {Biloshytskyi}}, \bibinfo {author} {\bibfnamefont {X.-L.}\ \bibnamefont {Ren}}, \ and\ \bibinfo {author} {\bibfnamefont {M.}~\bibnamefont {Vanderhaeghen}},\ }\href {\doibase 10.1103/PhysRevD.107.074021} {\bibfield  {journal} {\bibinfo  {journal} {Phys. Rev. D}\ }\textbf {\bibinfo {volume} {107}},\ \bibinfo {pages} {074021} (\bibinfo {year} {2023})},\ \Eprint {http://arxiv.org/abs/2206.15223} {arXiv:2206.15223 [hep-ph]} \BibitemShut {NoStop}%
\bibitem [{\citenamefont {Pelaez}(2016)}]{Pelaez:2015qba}%
  \BibitemOpen
  \bibfield  {author} {\bibinfo {author} {\bibfnamefont {J.~R.}\ \bibnamefont {Pelaez}},\ }\href {\doibase 10.1016/j.physrep.2016.09.001} {\bibfield  {journal} {\bibinfo  {journal} {Phys. Rept.}\ }\textbf {\bibinfo {volume} {658}},\ \bibinfo {pages} {1} (\bibinfo {year} {2016})},\ \Eprint {http://arxiv.org/abs/1510.00653} {arXiv:1510.00653 [hep-ph]} \BibitemShut {NoStop}%
\bibitem [{\citenamefont {Pel\'aez}\ and\ \citenamefont {Rodas}(2022)}]{Pelaez:2020gnd}%
  \BibitemOpen
  \bibfield  {author} {\bibinfo {author} {\bibfnamefont {J.~R.}\ \bibnamefont {Pel\'aez}}\ and\ \bibinfo {author} {\bibfnamefont {A.}~\bibnamefont {Rodas}},\ }\href {\doibase 10.1016/j.physrep.2022.03.004} {\bibfield  {journal} {\bibinfo  {journal} {Phys. Rept.}\ }\textbf {\bibinfo {volume} {969}},\ \bibinfo {pages} {1} (\bibinfo {year} {2022})},\ \Eprint {http://arxiv.org/abs/2010.11222} {arXiv:2010.11222 [hep-ph]} \BibitemShut {NoStop}%
\bibitem [{\citenamefont {Caprini}\ \emph {et~al.}(2006)\citenamefont {Caprini}, \citenamefont {Colangelo},\ and\ \citenamefont {Leutwyler}}]{Caprini:2005zr}%
  \BibitemOpen
  \bibfield  {author} {\bibinfo {author} {\bibfnamefont {I.}~\bibnamefont {Caprini}}, \bibinfo {author} {\bibfnamefont {G.}~\bibnamefont {Colangelo}}, \ and\ \bibinfo {author} {\bibfnamefont {H.}~\bibnamefont {Leutwyler}},\ }\href {\doibase 10.1103/PhysRevLett.96.132001} {\bibfield  {journal} {\bibinfo  {journal} {Phys. Rev. Lett.}\ }\textbf {\bibinfo {volume} {96}},\ \bibinfo {pages} {132001} (\bibinfo {year} {2006})},\ \Eprint {http://arxiv.org/abs/hep-ph/0512364} {arXiv:hep-ph/0512364} \BibitemShut {NoStop}%
\bibitem [{\citenamefont {Garcia-Martin}\ \emph {et~al.}(2011{\natexlab{b}})\citenamefont {Garcia-Martin}, \citenamefont {Kaminski}, \citenamefont {Pelaez},\ and\ \citenamefont {Ruiz~de Elvira}}]{Garcia-Martin:2011nna}%
  \BibitemOpen
  \bibfield  {author} {\bibinfo {author} {\bibfnamefont {R.}~\bibnamefont {Garcia-Martin}}, \bibinfo {author} {\bibfnamefont {R.}~\bibnamefont {Kaminski}}, \bibinfo {author} {\bibfnamefont {J.~R.}\ \bibnamefont {Pelaez}}, \ and\ \bibinfo {author} {\bibfnamefont {J.}~\bibnamefont {Ruiz~de Elvira}},\ }\href {\doibase 10.1103/PhysRevLett.107.072001} {\bibfield  {journal} {\bibinfo  {journal} {Phys. Rev. Lett.}\ }\textbf {\bibinfo {volume} {107}},\ \bibinfo {pages} {072001} (\bibinfo {year} {2011}{\natexlab{b}})},\ \Eprint {http://arxiv.org/abs/1107.1635} {arXiv:1107.1635 [hep-ph]} \BibitemShut {NoStop}%
\bibitem [{\citenamefont {Moussallam}(2011)}]{Moussallam:2011zg}%
  \BibitemOpen
  \bibfield  {author} {\bibinfo {author} {\bibfnamefont {B.}~\bibnamefont {Moussallam}},\ }\href {\doibase 10.1140/epjc/s10052-011-1814-z} {\bibfield  {journal} {\bibinfo  {journal} {Eur. Phys. J. C}\ }\textbf {\bibinfo {volume} {71}},\ \bibinfo {pages} {1814} (\bibinfo {year} {2011})},\ \Eprint {http://arxiv.org/abs/1110.6074} {arXiv:1110.6074 [hep-ph]} \BibitemShut {NoStop}%
\bibitem [{\citenamefont {Donoghue}\ \emph {et~al.}(1990)\citenamefont {Donoghue}, \citenamefont {Gasser},\ and\ \citenamefont {Leutwyler}}]{Donoghue:1990xh}%
  \BibitemOpen
  \bibfield  {author} {\bibinfo {author} {\bibfnamefont {J.~F.}\ \bibnamefont {Donoghue}}, \bibinfo {author} {\bibfnamefont {J.}~\bibnamefont {Gasser}}, \ and\ \bibinfo {author} {\bibfnamefont {H.}~\bibnamefont {Leutwyler}},\ }\href {\doibase 10.1016/0550-3213(90)90474-R} {\bibfield  {journal} {\bibinfo  {journal} {Nucl. Phys. B}\ }\textbf {\bibinfo {volume} {343}},\ \bibinfo {pages} {341} (\bibinfo {year} {1990})}\BibitemShut {NoStop}%
\bibitem [{\citenamefont {Moussallam}(2000)}]{Moussallam:1999aq}%
  \BibitemOpen
  \bibfield  {author} {\bibinfo {author} {\bibfnamefont {B.}~\bibnamefont {Moussallam}},\ }\href {\doibase 10.1007/s100520050738} {\bibfield  {journal} {\bibinfo  {journal} {Eur. Phys. J. C}\ }\textbf {\bibinfo {volume} {14}},\ \bibinfo {pages} {111} (\bibinfo {year} {2000})},\ \Eprint {http://arxiv.org/abs/hep-ph/9909292} {arXiv:hep-ph/9909292} \BibitemShut {NoStop}%
\bibitem [{\citenamefont {Hoferichter}\ \emph {et~al.}(2012)\citenamefont {Hoferichter}, \citenamefont {Ditsche}, \citenamefont {Kubis},\ and\ \citenamefont {Meissner}}]{Hoferichter:2012wf}%
  \BibitemOpen
  \bibfield  {author} {\bibinfo {author} {\bibfnamefont {M.}~\bibnamefont {Hoferichter}}, \bibinfo {author} {\bibfnamefont {C.}~\bibnamefont {Ditsche}}, \bibinfo {author} {\bibfnamefont {B.}~\bibnamefont {Kubis}}, \ and\ \bibinfo {author} {\bibfnamefont {U.~G.}\ \bibnamefont {Meissner}},\ }\href {\doibase 10.1007/JHEP06(2012)063} {\bibfield  {journal} {\bibinfo  {journal} {JHEP}\ }\textbf {\bibinfo {volume} {06}},\ \bibinfo {pages} {063} (\bibinfo {year} {2012})},\ \Eprint {http://arxiv.org/abs/1204.6251} {arXiv:1204.6251 [hep-ph]} \BibitemShut {NoStop}%
\bibitem [{\citenamefont {Blackstone}\ \emph {et~al.}(2024)\citenamefont {Blackstone}, \citenamefont {Tarr\'us~Castell\`a}, \citenamefont {Passemar},\ and\ \citenamefont {Zupan}}]{Blackstone:2024ouf}%
  \BibitemOpen
  \bibfield  {author} {\bibinfo {author} {\bibfnamefont {P.~J.}\ \bibnamefont {Blackstone}}, \bibinfo {author} {\bibfnamefont {J.}~\bibnamefont {Tarr\'us~Castell\`a}}, \bibinfo {author} {\bibfnamefont {E.}~\bibnamefont {Passemar}}, \ and\ \bibinfo {author} {\bibfnamefont {J.}~\bibnamefont {Zupan}},\ }\href@noop {} {\  (\bibinfo {year} {2024})},\ \Eprint {http://arxiv.org/abs/2407.13587} {arXiv:2407.13587 [hep-ph]} \BibitemShut {NoStop}%
\bibitem [{\citenamefont {Pich}\ \emph {et~al.}(2023)\citenamefont {Pich}, \citenamefont {Solomonidi},\ and\ \citenamefont {Vale~Silva}}]{Pich:2023kim}%
  \BibitemOpen
  \bibfield  {author} {\bibinfo {author} {\bibfnamefont {A.}~\bibnamefont {Pich}}, \bibinfo {author} {\bibfnamefont {E.}~\bibnamefont {Solomonidi}}, \ and\ \bibinfo {author} {\bibfnamefont {L.}~\bibnamefont {Vale~Silva}},\ }\href {\doibase 10.1103/PhysRevD.108.036026} {\bibfield  {journal} {\bibinfo  {journal} {Phys. Rev. D}\ }\textbf {\bibinfo {volume} {108}},\ \bibinfo {pages} {036026} (\bibinfo {year} {2023})},\ \Eprint {http://arxiv.org/abs/2305.11951} {arXiv:2305.11951 [hep-ph]} \BibitemShut {NoStop}%
\bibitem [{\citenamefont {Ablikim}\ \emph {et~al.}(2017{\natexlab{b}})\citenamefont {Ablikim} \emph {et~al.}}]{BESIII:2016bnd}%
  \BibitemOpen
  \bibfield  {author} {\bibinfo {author} {\bibfnamefont {M.}~\bibnamefont {Ablikim}} \emph {et~al.} (\bibinfo {collaboration} {BESIII}),\ }\href {\doibase 10.1103/PhysRevLett.118.092001} {\bibfield  {journal} {\bibinfo  {journal} {Phys. Rev. Lett.}\ }\textbf {\bibinfo {volume} {118}},\ \bibinfo {pages} {092001} (\bibinfo {year} {2017}{\natexlab{b}})},\ \Eprint {http://arxiv.org/abs/1611.01317} {arXiv:1611.01317 [hep-ex]} \BibitemShut {NoStop}%
\bibitem [{\citenamefont {Workman}\ \emph {et~al.}(2022)\citenamefont {Workman} \emph {et~al.}}]{ParticleDataGroup:2022pth}%
  \BibitemOpen
  \bibfield  {author} {\bibinfo {author} {\bibfnamefont {R.~L.}\ \bibnamefont {Workman}} \emph {et~al.} (\bibinfo {collaboration} {Particle Data Group}),\ }\href {\doibase 10.1093/ptep/ptac097} {\bibfield  {journal} {\bibinfo  {journal} {PTEP}\ }\textbf {\bibinfo {volume} {2022}},\ \bibinfo {pages} {083C01} (\bibinfo {year} {2022})}\BibitemShut {NoStop}%
\bibitem [{\citenamefont {Ablikim}\ \emph {et~al.}(2021)\citenamefont {Ablikim} \emph {et~al.}}]{BESIII:2020qkh}%
  \BibitemOpen
  \bibfield  {author} {\bibinfo {author} {\bibfnamefont {M.}~\bibnamefont {Ablikim}} \emph {et~al.} (\bibinfo {collaboration} {BESIII}),\ }\href {\doibase 10.1103/PhysRevLett.126.102001} {\bibfield  {journal} {\bibinfo  {journal} {Phys. Rev. Lett.}\ }\textbf {\bibinfo {volume} {126}},\ \bibinfo {pages} {102001} (\bibinfo {year} {2021})},\ \Eprint {http://arxiv.org/abs/2011.07855} {arXiv:2011.07855 [hep-ex]} \BibitemShut {NoStop}%
\end{thebibliography}%

\end{document}